\documentclass[a4paper,11pt]{article}
\usepackage{fullpage,hyperref}
\usepackage{cite}
\usepackage{fullpage}
\usepackage{lipsum}
\usepackage{amsmath}
\usepackage{amsfonts}
\usepackage{amssymb}
\usepackage{latexsym}
\usepackage{dsfont,eucal,mathrsfs}
\usepackage{graphicx}
\usepackage{subcaption}
\usepackage{bm}

\usepackage{ragged2e}
\usepackage{hyperref}
\DeclareCaptionJustification{justified}{\justifying}
\usepackage{todonotes}
\usepackage{changes}
\definechangesauthor[name={Deniz}, color=magenta]{DE}

\definechangesauthor[name={Matteo}, color=blue]{MT}

\definechangesauthor[name={SebMat}, color=red]{S}

\definechangesauthor[name={Tiago}, color=blue]{tp}

\definechangesauthor[name={Methods}, color=blue]{Methods}

\newcommand{\spacing}[1]{\renewcommand{\baselinestretch}{#1}\large\normalsize}
\spacing{1}

\newenvironment{affiliations}{%
    \setcounter{enumi}{1}%
    \setlength{\parindent}{0in}%
    \slshape\sloppy%
    \begin{list}{\upshape$^{\arabic{enumi}}$}{%
        \usecounter{enumi}%
        \setlength{\leftmargin}{0in}%
        \setlength{\topsep}{0in}%
        \setlength{\labelsep}{0in}%
        \setlength{\labelwidth}{0in}%
        \setlength{\listparindent}{0in}%
        \setlength{\itemsep}{0ex}%
        \setlength{\parsep}{0in}%
        }
    }{\end{list}\par\vspace{12pt}}

\usepackage{mwe}

\newcommand{\bo}{\boldsymbol}

\renewcommand{\mathbf}{\boldsymbol}
\renewcommand{\mathcal}{\mathscr}

\newtheorem{theorem}{Theorem}

\title{Effective networks: a model to predict network structure and critical transitions from datasets
}
\author
{Deniz Eroglu,$^{1,2,3}$ Matteo Tanzi,$^{2,4}$ Sebastian van Strien,$^2$ Tiago Pereira$^{1,2}$}

\date{}

\begin{document}

\maketitle

\begin{affiliations}
 \item Instituto de Ci\^encias Matem\'aticas e Computa\c{c}\~ao, Universidade de S\~ao Paulo, S\~ao Carlos, Brazil
 \item Department of Mathematics, Imperial College London, London SW7 2AZ, UK
  \item Department of Bioinformatics and Genetics, Kadir Has University, 34083 Istanbul, Turkey
 \item Department of Mathematics and Statistics, University of Victoria, Victoria BC, CA
 \end{affiliations}

\begin{abstract}
Real-world  complex systems such as ecological communities and neuron networks are essential parts of our everyday lives.  These systems are composed of units which interact through intricate networks. The ability to predict sudden changes in network behaviour, known as critical transitions, from data is important to avert disastrous consequences of major disruptions. Predicting such changes is a major challenge as it requires forecasting the behaviour for parameter ranges for which no data on the system is available. In this paper, we address this issue for  networks with weak individual interactions and chaotic local dynamics. We do this by building a model network, termed an {\em effective network}, consisting of the underlying local dynamics at each node  and a statistical description of their interactions. We illustrate this approach by reconstructing the dynamics and structure of  realistic neuronal interaction networks of the cat cerebral cortex.  We reconstruct the community structure by analysing the stochastic fluctuations generated by the network  and predict critical transitions for coupling parameters outside the observed range.
\end{abstract}  
 \vspace{5mm}





We are surrounded by a range of complex networks consisting of a large number of intricately coupled nodes.  Neuron networks form an important class examples where the interaction structure is heterogeneous \cite{kandel2000}. Because changes in the interaction can have massive ramifications on the system as a whole, it is desirable to predict such disturbances and thus enact precautionary measures to avert potential disasters.  For instance, neurological disorders such as Parkinson's disease, schizophrenia, and epilepsy, are thought to be associated with anomalous interaction structure among neurons \cite{bohland2009}. Often, as in the case of neuron networks, it is impossible to directly determine the interaction structure.  Therefore, a major scientific challenge is to develop techniques which use measurements of the time evolution of the nodes state to indirectly recover the network structure and predict the network's behaviour when essential characteristics of the interactions change. 
 
The literature on data-based network reconstruction is vast. Reconstruction methods can be classified into {\it model-free} methods and \emph{model-based} methods.  The former identify the presence and strength of a connection between two nodes  by measuring the {\it dependence} between their time-series in terms of:  correlations \cite{de2004discovery,reverter2008combining}, mutual information \cite{butte1999mutual}, maximum entropy distributions \cite{braunstein2008inference,cocco2009neuronal}, Granger causality, and causation entropy  \cite{bressler2011wiener,ladroue2009beyond}. Such methods alone do not provide information on the dynamics of the network, which is necessary to predict critical transitions. Model-based methods instead provide estimates (or assume a priori knowledge) of the dynamics and interactions in the system, and use this knowledge to reconstruct the network topology. When the interactions are sufficiently strong, the network structure can be recovered \cite{casadiego2017,han2015,wang2016}.  For a more extensive account of reconstruction (model-free and -based) methods see the reviews \cite{wang2016,nitzan2017revealing,stankovski2017} and references therein. 

In many real-world applications, the behaviour of isolated nodes is erratic (\emph{chaotic}) and the interaction is weak \cite{schneidman2006,haas2015,kandel2000}. Moreover, the network structure typically has community structures and hierarchical organisations such as the rich-club networks found in the brain \cite{heuvel2011}. As the interaction strength per connection is weak and the statistical behaviour of the nodes robust, the influence of a single link on the overall dynamics is negligible and only the cumulative contribution of many links matter. Furthermore, because of the chaotic dynamics and weak coupling, the influence of  a single node on the rest of the network corresponds  essentially to a random signal that is superposed to the randomness generated by the chaos in the local dynamics. The existing reconstruction techniques fail to reconstruct a model from the data, as they require the interaction to be of the same magnitude as the isolated dynamics. 
 
In this paper, we introduce the notion of an {\em effective network}. Its aim is to make a model of a complex system from observations of the nodes evolution when the network has heterogeneous structure,  the strength of interaction is small and  local dynamics are highly erratic. This approach starts by reconstructing the local dynamics from observations of nodes with relatively few connections, and then recovers the interaction function from observations of the highly connected nodes whose dynamics are the most affected by the interactions as a result of the multitude of connections they receive from the rest of the networks\cite{park2013,pereira2017}. A key achievement is that we are able to identify  community structures even in the presence of weak coupling.  An outcome of the effective network  is that it recovers enough information to forecast  and  anticipate the network behaviour,  even in situations where the parameters of the system change into ranges that have not been previously encountered. 
 
\noindent 
{\bf Complex networks of nonlinear systems.} We consider  networks with $N$ nodes with chaotic isolated dynamics and pairwise interactions satisfying some additional assumptions described below. The network is described by its adjacency matrix  $\bo A$, whose entry $A_{ij}$ equals $1$ if node $i$ receives a connection from $j$ and equals $0$ otherwise. We assume that the  time evolution of the state $\bo x_i(t)$ of node $i$ at time $t$ is expressed  as
\begin{eqnarray}
\bm{x}_i(t+1)&=&\bm{F}_i (\bm{x}_i(t))+\alpha \sum_{j=1}^{N} A_{ij} \bm{H}(\bm{x}_i(t),\bm{x}_j(t)). 
\label{eq1}
\end{eqnarray} 
When performing reconstruction from data, the isolated local dynamics $\bm{F}_i\colon M\to M$, the coupling function $\bm{H}$, the coupling parameter $\alpha$ (that is small), the adjacency matrix $\bo A$,  and even the dimension $k$ of the space $M$, are assumed to be  {\em unknown}. This class of equations can be applied to modelling many important complex systems found in neuroscience (e.g. coupled equations in neuron networks) \cite{izhikevich2007}, engineering (e.g. smart grids) \cite{yadav2017,dorfler2013}, material sciences (e.g. superconductors) \cite{watanabe1994}, and biology (e.g. cardiac pacemaker cells) \cite{winfree2001}, among other systems.  

Our three assumptions are: $(a)$ the local dynamics are close to some unknown {\em ergodic} and {\em chaotic} map $\bm{F}$ (that is, $\| \bm{F} - \bm{F}_i \| \le \delta$, which is often the case in applications \cite{pinto2000,eroglu2017}). $(b)$ The network connectivity is {\em heterogeneous}, which means that the  number of incoming connections at a node  $i$ (given by its degree $k_i = \sum_j A_{ij}$) varies widely across the network.  More precisely, $k_i$ is larger for a few nodes called {\em hubs}. $(c)$ $\alpha$ is such that, denoting by $\Delta=\max_i k_i$ the maximum number of connections  $\alpha \Delta$ is of the same magnitude of $\bm{DF}$.  Assumptions $(a)$ and $(c)$ imply that the the total effect of the network on a node $i$, which is of the order $\alpha k_i$, ranges from small (for nodes $i$ with few incoming connections) to of order 1 for the hub nodes. A prime example  is the cat cerebral cortex which possesses inter-connected regions split into communities with a hierarchical organization. This network has heterogeneous connectivity, chaotic motion and weak coupling \cite{scannell1993,scannell1995,zamora-lopez2010}. Other examples, includes the drosophila optic lobe network \cite{takemura2013,garcia-perez2018}. For a given dataset, our effective network first tests whether the underlying system satisfies assumptions $(a)-(c)$, and, if so, reconstructs the model (see the Methods for more detail).

We assume the availability of  a time series of  observations $y_i(t)=\phi( {\bo x}_i(t))$,  where $\phi$ is a projection to a variable on which unit interactions depend. This situation occurs frequently in applications; for example, with  measurements of membrane potentials in neurons or voltages in electronic circuits. Here,  we demonstrate how to obtain a model for the system by constructing an effective network from the time series $y_i(t)$.
 
\section*{Effective networks recover both structure and dynamics}

To obtain an {\em effective network} from observations of a complex system, we combine statistical analysis, machine-learning techniques, and dynamical systems theory for networks. An effective network provides (i)  local evolution laws and averaged interactions for each unit that, in combination, closely approximate  the unit dynamics, and (ii) a network with the same degree distribution and communities  as the original system. We use the term ``effective" because, even if it does not carry information on each link and interaction in the original system,  the network gathers sufficient data to reproduce the behaviour of the original network and predict its critical transitions.

\begin{figure}[h]
    \centering
    \begin{subfigure}[t]{1.0\textwidth}
        \centering
        \includegraphics[width=\linewidth]{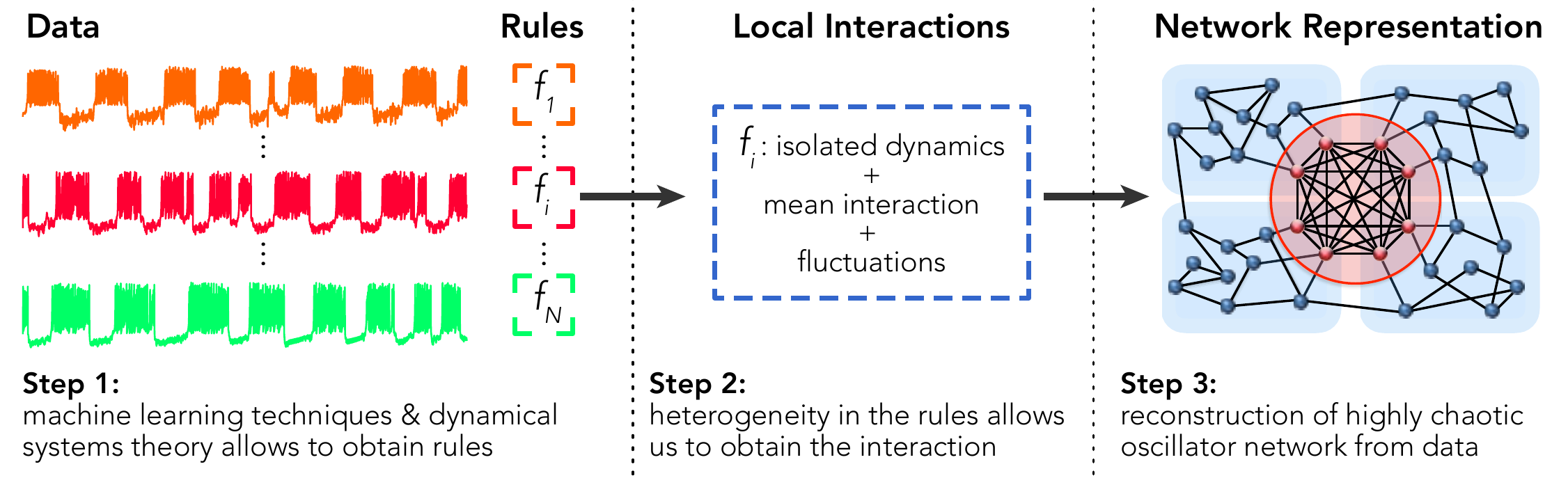}
    \end{subfigure}%
\caption{{\bf Reconstruction scheme with {the effective network.}}  From the time series, we build a model for the local evolution  $f_i$ at each node. Under the assumption that such rules change from node to node depending on their connectivity, we estimate the coupling function. Using the fluctuations of the time series with respect to the low-dimensional rules, we recover the community structures. Gathering all this information, we obtain an effective network which is a model for the original system that can be used to predict critical transitions}.
\label{RC} 
\end{figure}

Using  our assumptions for the network and local dynamics, we can show that the evolution at each node will have low-dimensional excursions over finite time scales. In particular, the approximated evolution rule  at node $i$ has a low dimensional description over large time scales given by . More precisely, the evolution rule at node $i$ is given by
\begin{eqnarray}
\bo G_i(\bo x_i)  = \bo F_i(\bo x_i) + \beta_i \bo V(\bo x_i(t)) \nonumber
\end{eqnarray}
where $\bo F_i \approx \bo F$  is the isolated dynamics, $\beta_i = \alpha k_i$ is the rescaled degree, and $\bo V$ is the averaged mean-field effect of the interaction function (effective coupling function) that takes into account the cumulative effect of interactions on node $i$.  The true dynamics $\bo x_i(t)$ slightly deviates from this rule and is given by 
\begin{eqnarray}
\bo x_i(t+1) = \bo G_i(\bo x_i(t)) + \bo \xi_i(t) \nonumber
\end{eqnarray}
where $\bo \xi_i(t)$ is a fluctuation term that is  small for an interval of time which is exponentially large in terms of the size of the network.
This low-dimensional reduction  has been rigorously established in important test cases over long time scales (see \cite{pereira2017}). 
%
This approximation holds, roughly speaking, for two reasons. Firstly, a low degree node $i$ has a  small number of  connections compared to the maximum degree, and the interactions slightly perturb its statistical behaviour.  Secondly,  hub nodes receive a huge number of connections so that the sum of their interactions with the other nodes in the network, $\alpha \sum_j A_{ij} \bm{H}(\bm{x}_i,\bm{x}_j)$, can be approximated as the integral $\alpha k_i \int \bm{H}(\bm{x}_i,\bm{y})\rho (\bm{y}) \, d\bm{y}$ where $\rho(\bm{y})$ is the stationary distribution of typical orbits of the low degree nodes. This approximation holds up to a small fluctuation, $\bo \xi(t)$, which depends on the state of neighbours of the $i$th node. This fact will be  key to detecting communities.
%

The approximation described above also applies to the measured state variable $y_i(t)$. Depending on the system, we need to preprocess the data (see  Methods). The processed variable is still referred to as $y_i(t)$. The Takens reconstruction technique tells us that  $y_i(t+1)$  is a nonlinear function of $k+1$ past points $y_i(t), \dots y_i(t-k)$, for a given number $k$ provided by the approach. Here, we focus on the case when $k=1$, which occurs in many real-world examples, and discuss cases with $k\ge2$ in the Supplementary Materials. This means that 
\begin{eqnarray}
y_i(t+1) = g_i(y(t))+ \xi_i(t) \label{eq:meanfield}
\end{eqnarray}
where 
\[g_i=f_i(y_i(t))+\beta_i v(y_i(t))
\] is a low-dimensional law that approximates the evolution of $y_i(t)$.
We first employ Takens reconstruction to estimate  $g_i$ and coarsely classify the nodes by their degrees. For instance, in rich-club motifs, the network has low-degree nodes organised in communities and high-degree in the rich-club. Since $\beta_i=\alpha k_i$ is small for low-degree nodes $i$,  the evolution rules at such nodes will be similar to the isolated dynamics. Thus we identify 
which nodes are low-degree nodes and we can recover the local dynamics $f$.  Next, we use Eq. (\ref{eq:meanfield}) and a classification
of nodes by their time-series, to obtain the coupling function and estimate the degree distribution $k_i$ and the coupling parameter $\alpha$. 
More precisely, an effective network is obtained in three main steps (see Methods for additional details): 

\begin{itemize}

\item[(i)] {\bf Reduced dynamics.}  
Since the fluctuation term $\xi_i(t)$ is small,  the rule $g_i$ can be estimated fitting the points $(y_i(t), y_i(t+1))$  using machine-learning techniques. Here, we estimate the function $g_i$ by decomposing it as a linear combination of basis functions, which is chosen according to the application. The parameters of the basis functions are obtained by performing a $10$-fold cross-validation with $90\%$ training and $10\%$ test \cite{shandilya2011, james2013} (see Methods for details). The particular technique used in this step is not essential as the dynamics will be low-dimensional. Other techniques such as compressive sensing based techniques \cite{brunton2015, mangan2016,wang2011} could be also employed.  

\item[(ii)]  {\bf Isolated dynamics and effective coupling.} Since the network is heterogeneous, it has many low degree nodes for which $g_i$ is  close to the uncoupled dynamics $f_i\approx f$. We first run a model-free estimation to coarsely classify nodes into low-degree nodes and hubs and estimate $f$ (see Supplementary Material).  We then use the expression of $g_i$ recovered at low degree nodes to obtain an approximation for $f$, while $g_i$ recovered at hub nodes allows to estimate $\beta_i v$.  We estimate the parameter $\beta_i$ by using Bayesian inference (see Methods).

\item[(iii)] {\bf Network structure and communities.} Since $\beta_i=\alpha k_i$, we can recover the network's degree distribution from knowledge of $\beta_i$ at every node. To reconstruct the community structures, we use the fact that the term $\xi_i(t)$ at a node $i$ depends on the state of all nodes interacting with $i$. 
Thus the correlation between $\xi_i$ and $\xi_j$ is proportional to  the matching index (number of common neighbors) of the nodes $i$ and $j$. Thus, we can create a network that has the same statistical properties of the actual one. Further details are provided in the next section.

 \end{itemize}
 
 \noindent
{\bf Benchmark model for the isolated dynamics.}  We consider heterogeneous networks of neurons described by the  Rulkov model (see Methods for details). The model has two variables, $u$ and $w$, evolving at different time scales. The fast variable $u$ describes the membrane potential and $w$ the slow currents. 
In the following discussion, we choose the parameters such that the isolated dynamics presents a tonic spiking behaviour. In the Supplementary Materials, we also show the results for bursting dynamics, i.e.  when the parameters are such that the membrane potential oscillates between two phases: a bursting phase of a rapid spiking, and a quiescent phase. We also validated our methods for other types of isolate chaotic dynamics (also in the Supplementary materials). 

\section*{Revealing community structure: the rich-club motif}

We apply the effective network in a wide range of setups as discussed in the supplementary material. Here we focus on the network structure of the cat cerebral cortex for which the mesoscopic connectivity  has been probed and detailed \cite{zamora-lopez2010}.  
The regions and their connections were discovered by using datasets from tract-tracing experiments analyzed  via  nonmetric multidimensional scaling, an optimization approach that minimizes the distance between connected structures and maximizes  between the unconnected ones \cite{scannell1993,scannell1995}.

\begin{figure}[h]
    \centering
    \begin{subfigure}[t]{1\textwidth}
        \centering
        \includegraphics[width=\linewidth]{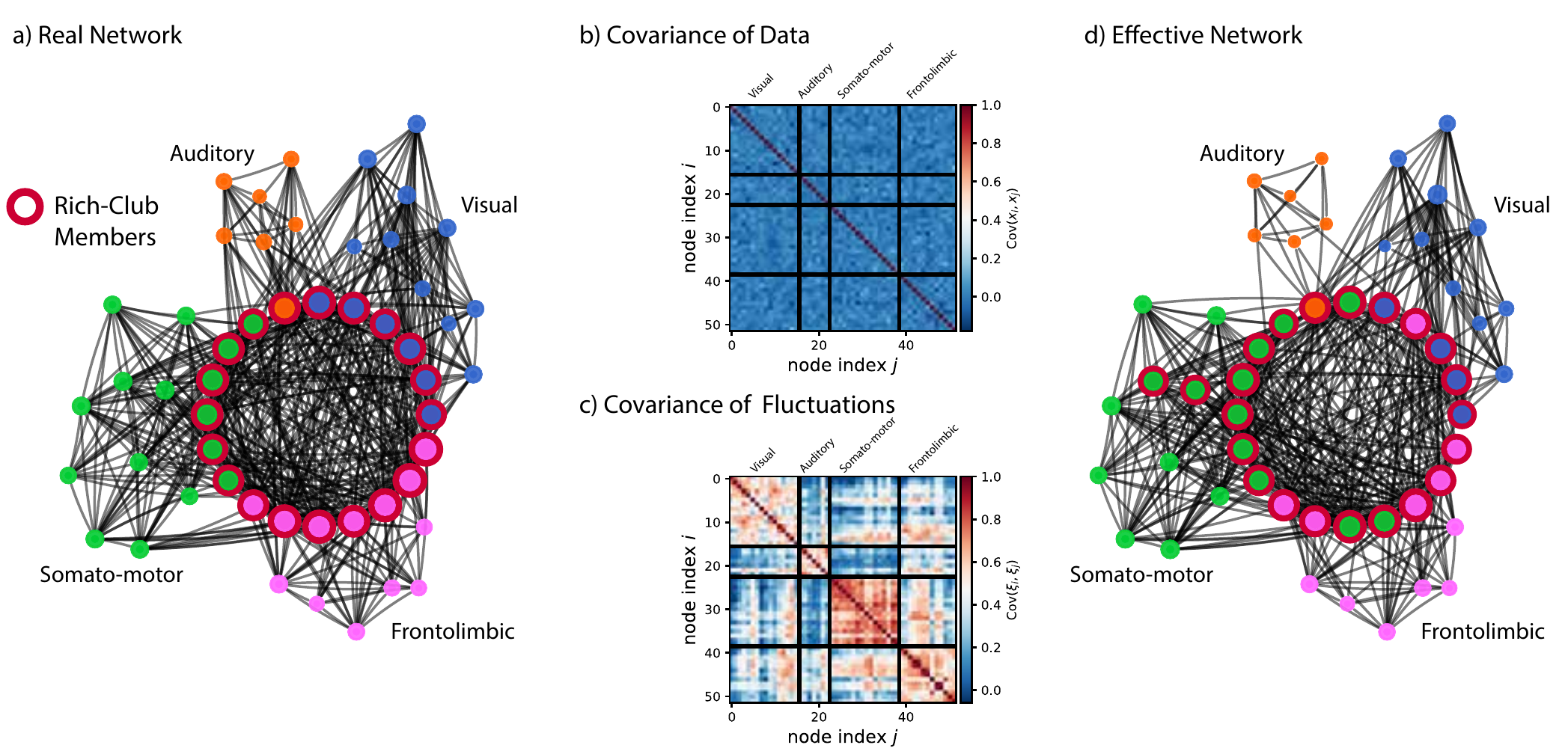}
    \end{subfigure}%
\caption{{\bf Effective network of the cat cerebral cortex.}  We use the local dynamics as a spiking neuron coupled via electric synapses. (a) The cat cerebral cortex network with nodes colour coded according to the four functional modules. Rich-club members are indicated by  red encircled nodes. (b) \emph{The covariance matrix of the data} cannot detect communities. (c) \emph{The covariance matrix of the fluctuations} can distinguish clusters of interconnected nodes. This matrix has entries color coded according to the key on the right with red entries corresponding to couple of nodes sharing a large numbers of nearest neighbours in the network, while blue nodes correspond to couple of nodes that share a  small number of common neighbours.   (d) A model in the cat cortex constructed via the effective network approach. From the matrix in (c) we can recover a representative effective network.  The reconstructed network represents the real network in (a) with good accuracy. See Methods for the details of the detection of communities and rich-club members.}
\label{RC2} 
\end{figure}

The network contains 53 meso-regions arranged in  four communities that closely follow functional subdivisions; namely,  visual (16 nodes), auditory (7 nodes), somatomotor (16 nodes) and frontolimbic (14 nodes), as shown in Fig \ref{RC2} (a). In addition to these cortical regions, some cortical areas (hubs) form a hidden layer called a \emph{rich-club} and are densely connected to each other and the communities. A set of nodes form a rich-club if their level of connectivity exceeds what would be expected by chance alone. The maximum number of connections in this network is $\Delta = 37$.
The network obtained is weighted \cite{scannell1993,scannell1995}. For simplicity  and to improve the performance  in detecting communities, we turn the network into an undirected simple graph \cite{zamora-lopez2010}.
%
We simulate each mesoregion as a neuron interacting via electrical synapses (see Methods for details) and obtain a multivariate data  $\{ y_1(t), y_2(t), \dots, y_N(t) \}$ for a time $T=5000$.
For simplicity, we will denote $y_i = \{ y_i(t)\}_{t=0}^T$, whenever there is no risk of confusion.

For comparison, we first test the recovery of the network using the functional network framework \cite{bettinardi2017,eguiluz2005,bullmore2009}. The intuition behind  this approach is that nodes with similar time series perform analogous functions in the network and are assumed to have similar characteristics. The functional network can thus be constructed by the matrix of the similarities between nodes via statistical analysis \cite{zhang2006,greicius2003}. Here we employ a covariance analysis between the time series. The functional network alone does not detect communities from the data because the bursts are erratic  (Fig. \ref{RC2} (b)). Other similarity measures such as mutual information give no significative improvement.

We also compared our method with a  reconstruction by the sparse recovery technique \cite{brunton2015,wang2016}.  There, the dynamics is written as linear combination of chosen  basis functions and the unknowns are the coefficients.  Sparse recovery works well when many coefficients are exactly zero. The presence of a link is determined when any coefficient of the corresponding interaction is nonzero.  In our setting, each link provides a small contribution  to the individual dynamics and only their cumulative effect can influence the dynamics.  Therefore, the coefficients to be recovered are close to zero, and cannot be distinguished from zero terms. In this setup, sparse recovery can also uncover a model for the isolated dynamics, however, it misidentifies the network structure. We implemented the sparse recovery method to our benchmark method and could assign only $20\%$ of the nodes to the right cluster. These results are discussed in the Methods. 

Remarkably, however, the effective network is able to recover the community structures (Fig. \ref{RC2} (c)). To  every pair of $y_i$ and  $y_j$ we assign  a \emph{Pearson distance} $s_{ij}\ge0$ which measures the difference between the observed dynamics at nodes $i$ and $j$ so that $s_{ij} \approx 0$ if the attractors of $i$ and $j$ are similar  and $s_{ij} \approx 1$ if they are reasonably distinguishable (see Methods for details). The higher the number of nodes with behaviour different from $i$, the larger  the intensity  $S_{i} = \sum_{j}s_{ij}$. The analysis of intensities $S_i$ allows us to distinguish nodes in the rich-club. Next, we find the local low-dimensional rules $g_i$ for each time series $y_i$ by applying  Step (i). These rules are  similar and display a chaotic evolution, thus, confirming that the similarity analysis by the functional networks would provide little insight into the reconstruction.
Having the local rules $g_i$, we can decompose the time series in terms of a low-dimensional deterministic part and a fluctuation term $\xi_i$, which depends on the neighbours of the $i$th node. One of our main key points is that these fluctuations allow us to reveal the community structure.  

Indeed, if nodes $i$ and $j$ interact with the same nodes, they are subject to the same inputs and  the covariance Cov$(\xi_j,\xi_i)$ is high. If not,  Cov$(\xi_i,\xi_j)$ is nearly zero due to the decay of correlations in the deterministic part. Therefore, Cov$(\xi_j,\xi_i)$ is high when nodes $i$ and $j$ have high {\em matching index} (high fraction of common connections). It is crucial that the correlation analysis is restricted to  the dynamical fluctuations $\xi_i$. Indeed, since the variance of the deterministic part of $y_{i}$ is larger than that of the small fluctuations $\xi_i$, performing a direct correlation analysis between $y_{i}$ and $y_{j}$ hides all the contributions to the correlations coming from the covariance between $\xi_i$ and $\xi_j$. Consequently, the correlation of the deterministic part  is close to zero due to the chaotic nature of the dynamics (see Supplementary Materials).  
So, remarkably, the noise terms $\xi_i$ are needed to access the community structure.

Using steps (i)-(iii) we obtained a model for the isolated dynamics, coupling function, and distribution of degrees.  To apply current methods of community detection, we obtain an adjacency matrix  with entries equal to zero or one from the matrix of correlations Fig. \ref{RC} (c). This is done by thresholding the entries and considering a link only when the correlations were greater than $0.5$. We tested threshold values ranging from  $0.3$ to $0.6$ and obtained the same results. Indeed, the distribution of the entries of the matrix of correlations is unimodal and has a peak near 0.5.  We computed rich-club coefficients for each node \cite{colizza2006} (see details in Methods), and nodes with  the highest coefficients formed the rich-club  displayed at the centre of the network. By applying the community detection method \cite{blondel2008} to the thresholded effective matrix, we recovered the communities according to their function as shown in Figure   \ref{RC2} (d). 


There is no reason to expect that the links in the effective network  correspond to links in the actual network. However, since every node makes most of its interactions within a cluster,  two nodes with highly correlated fluctuations $\xi(t)$ are likely to belong to the same community, and this can be enforced in the effective network by adding a connection between them.
 Further results for a variety of rich-club motifs are shown in the Supplementary Materials.

\paragraph{Performance of the communities reconstruction.} To quantify the effectiveness of  community reconstruction, we compute  $PE= \frac{m}{N}$, where $N$ is the total number of nodes and $m$ is the number of  nodes assigned to the wrong community. 
%
We compute $PE$ for different values for the coupling strength  $\Delta \alpha$ between $0.05$ and $0.4$. 
%
For each value of $\alpha$, we considered 50 different simulations of the dynamics obtained by choosing different initial conditions. 
The figure shows the plot of the mean of $PE$ and a shaded region corresponding to the standard deviation. For $\Delta \alpha$ values larger than $0.4$, the reconstructing procedure cannot identify the communities correctly. This is due to the synchronization rich club which appears around this value.

\begin{figure}[h]
    \centering
    \begin{subfigure}[t]{0.8\textwidth}
        \centering
        \includegraphics[width=\linewidth]{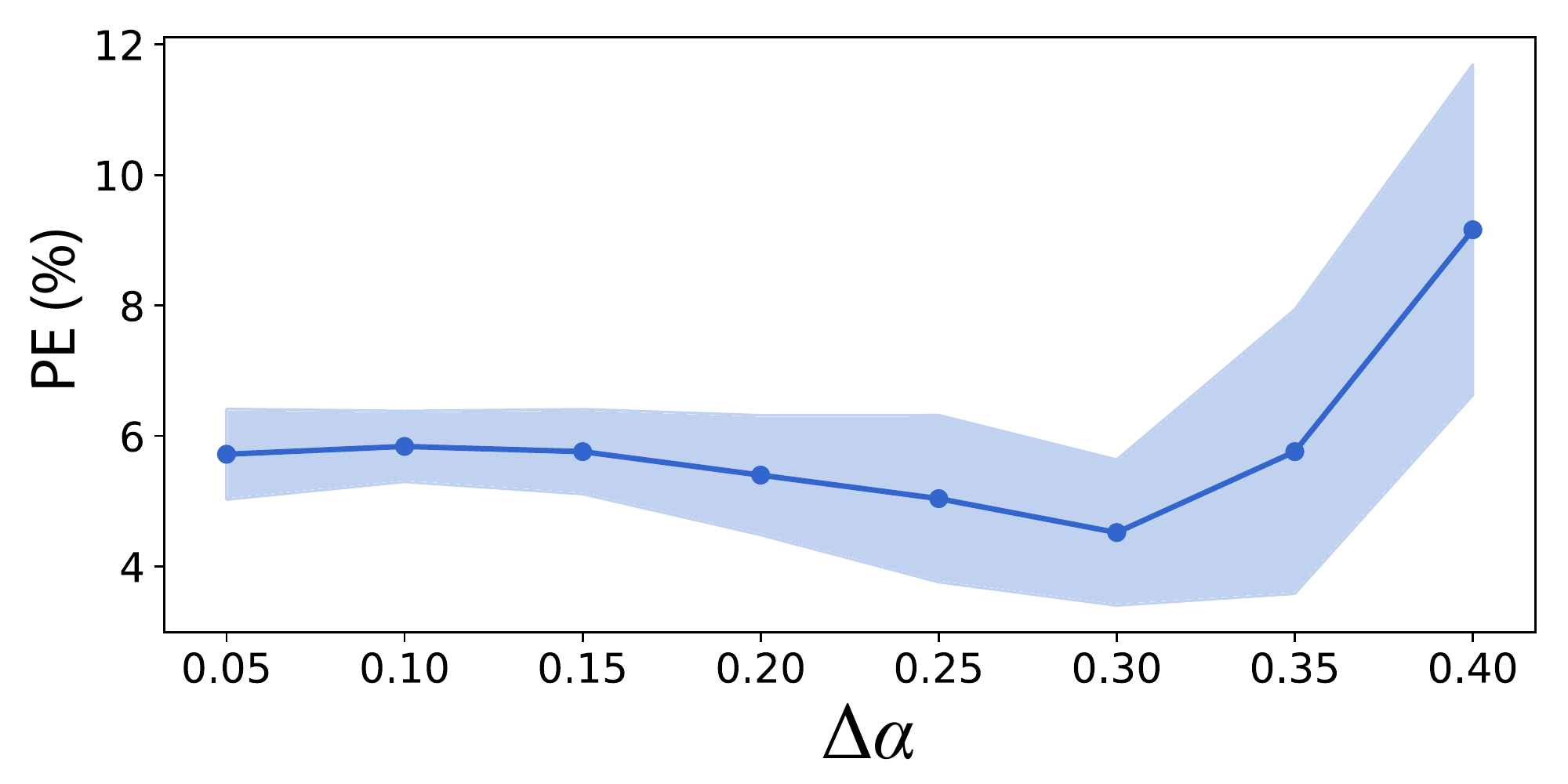}
    \end{subfigure}%
\caption{{\bf Prediction error for misidentification of communities in the reconstructed cat cerebral cortex from synthetic data.} For each realisation, the chosen parameters are the same as in Figure \ref{RC2} and only the overall coupling is changed. Mean and standard deviation of prediction error (PE) computed for the network over 50 realizations for each value of $\alpha$. If $\Delta \alpha > 0.42$, the system synchronizes and the procedure cannot reconstruct the community structures.}
\label{fig:pe} 
\end{figure}

\section*{Predicting critical transitions in rich-clubs}

The ability to reconstruct the network and dynamics from data can be exploited to predict critical transitions that may occur when the coupling strength. In the cat brain, for example, a transition to collective dynamics in the rich-club has drastic repercussions for the functionality of the network \cite{zamora-lopez2010,lopes2017}. 

Our goal now is to obtain data when the network is far from a collective dynamics and predict the onset of collective motion in the rich-club.  The effective network can predict the onset of such collective dynamics based on a single multivariate time series for fixed coupling strength in a regime far from the synchronized state. 
%
We analyze time-series obtained simulating the dynamics  at a value of the coupling strength $\Delta \alpha = 0.3$,
%
  and apply our reconstruction procedure to obtain the network structure and a model of the isolated dynamics.  

{\it Estimating the transition to synchronization in the rich-club.}  Transitions to synchronization between the bursts is possible while the fast spikes remain out of synchronization \cite{rulkov2001}. To estimate the transition to burst synchronization, we obtain the slow variable as a filter over the membrane potential (fast variable). Indeed, since we measure the membrane potential $y_i(t) = u_i(t)$, the slow variable is given as
$
z_i(t) = \mu \sum_{k=1}^t (y_i(k) - \sigma)
$
 and for a good choice of parameters $\mu$ and $\sigma$ this can be identified with the slow variable of the model $w$. In the methods section, we obtain an equation for the slow variable and analyze the effect of the network of the dynamics of the slow variable. 

We can use the data on the network and the dynamics recovered from the time-series recorded at  $\Delta \alpha = 0.3$  to predict that at the value $\Delta \alpha \approx 0.42$ the rich-club will develop a burst synchronization (details in the Methods).   %

\begin{figure}[h]
    \centering
    \begin{subfigure}[t]{0.9\textwidth}
        \centering
        \includegraphics[width=0.9\linewidth]{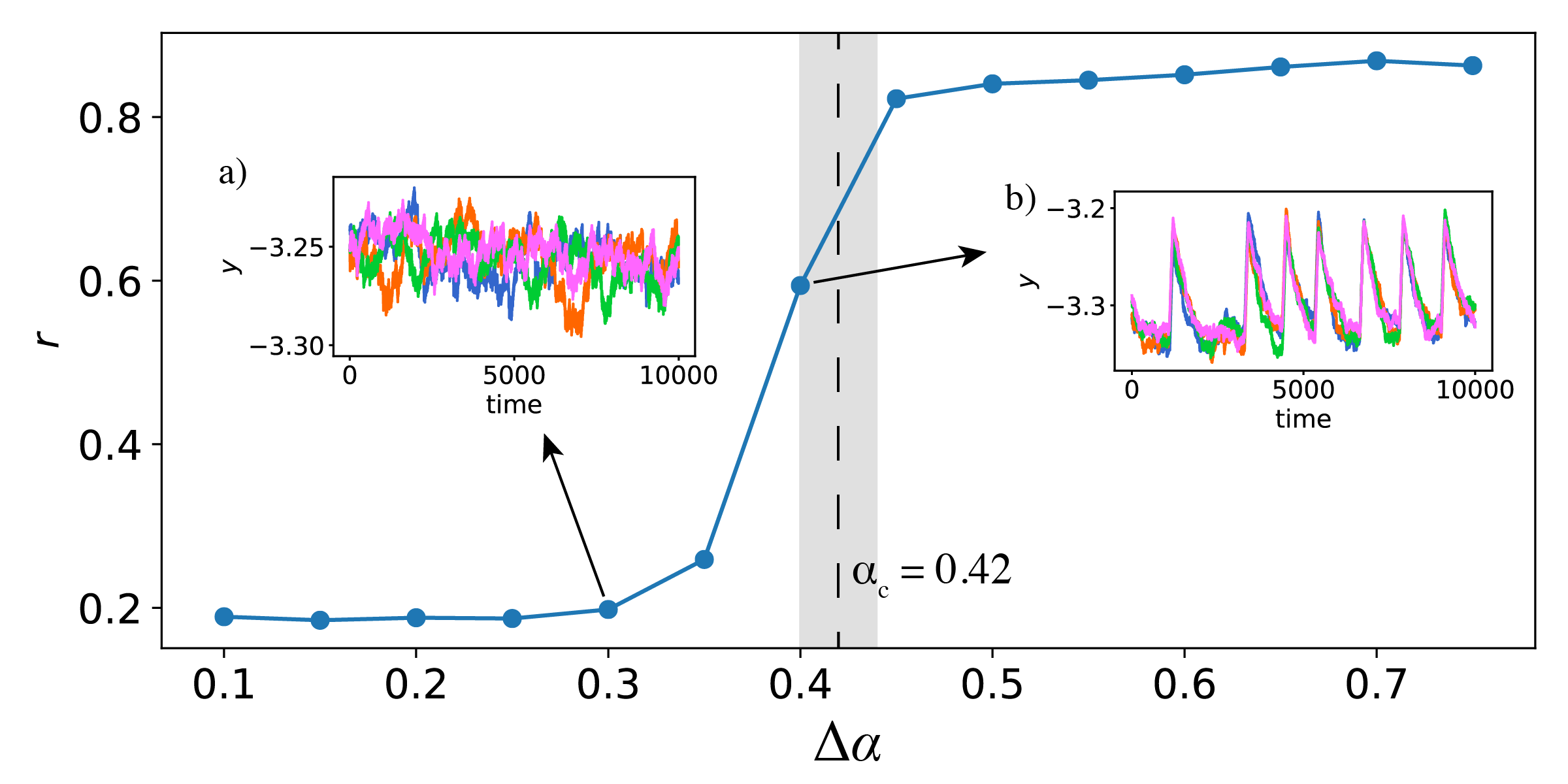}
\end{subfigure}%
\caption{{\bf Prediction of critical transitions in the rich-club of the cat cerebral cortex.} The level of average synchronization $r$ of the rich-club is shown for different values of the coupling strength. Insets show time series of neuronal dynamics of four rich-club members and color of time series matches with the color of nodes in Fig.~\ref{RC2}. For values in the grey shaded region, $r$ is increasing towards close to one and the rich-club exhibits collective behavior. We can predict the critical coupling $\alpha_c$ with some standard deviation (grey shaded region)  by studying the effective network obtained from a time series measured when $\Delta \alpha=0.3$.}
\label{fig:criticality} 
\end{figure}

To capture such a transition in the bursts of the rich club, we introduce a phase $\theta_j(t)$ for the slow variable, the definition of $\theta_j(t)$ can be found in the Methods section. Once we have the phase we compute the order parameter 
\begin{equation*}
r(t) e^{i \psi(t)} = \frac{1}{N_c} \sum_{j=1}^{N_c} e^{i \theta_j (t)}
\end{equation*}
Loosely speaking,  a small value of the order parameter, $r \approx 0$, means that no collective state is present in the system, whereas $r(t)\approx 1$ means that the bursts are synchronized.   Figure \ref{fig:criticality} shows that behaviour of the order parameter as a function of the coupling. 
%
The rich-club undergoes a transition to burst synchronization at $\Delta \alpha \approx 0.4$ that corresponds to an increase of roughly $40\%$ of the coupling strength and is close to the predicted value $\Delta \alpha \approx 0.42$.  
In the Supplementary Materials, we present other examples where the local dynamics is fully chaotic. 

%

\section*{Obtaining  a statistical description of the network}

The effective network can also provide  detailed information on the structure of each community and a statistical description of the network. To illustrate this, we reconstruct the statistical properties of scale-free networks.

\begin{figure}[t]
      \centering
    \begin{subfigure}[t]{\textwidth}
        \centering
        \includegraphics[width=\linewidth]{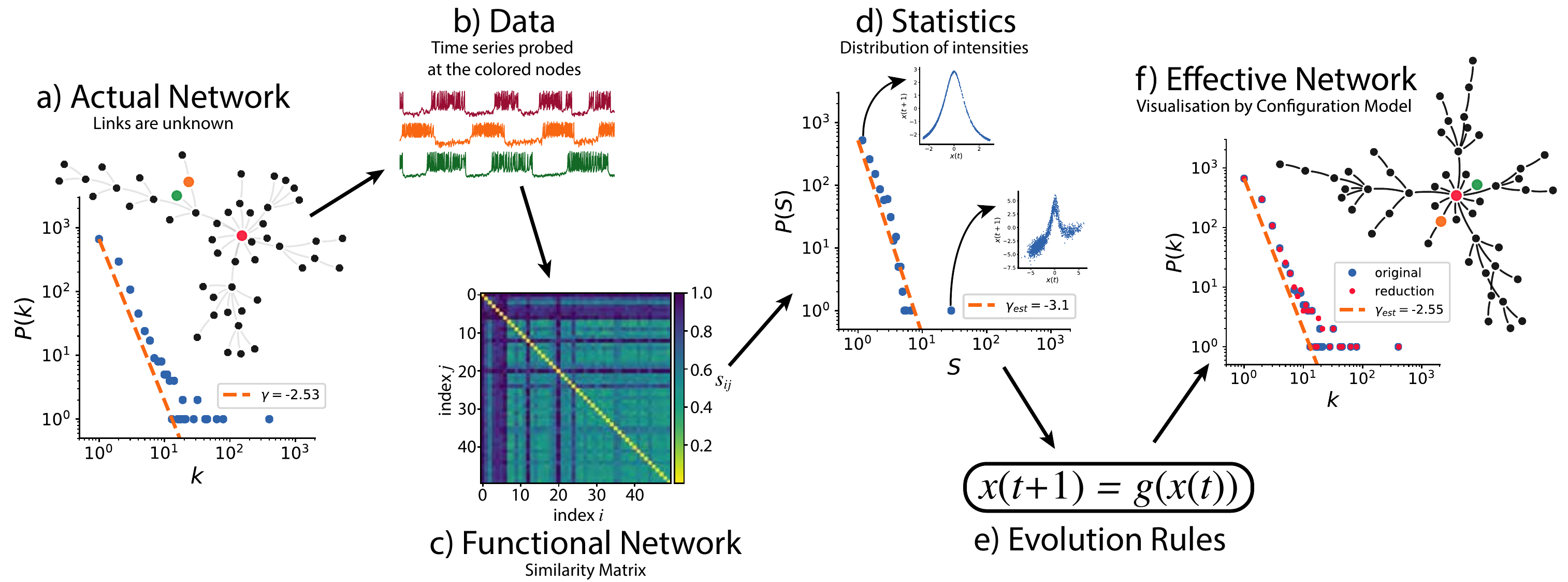}
    \end{subfigure}%
\caption{  
 {\bf Example of an effective network for a scale-free network where the interaction dynamics and  structure are unknown.} An effective network is constructed starting from a time series   and provides models for the network connectivity and unit  dynamics and interactions. The actual network and its degree distribution are shown in panel (a).  The time series of some representative nodes  are shown in panel (b). Correlation analysis of the time series yields the functional network represented in panel (c). Here,  the $(i,j)$ entry of the matrix is  the Pearson distance $s_{ij}$ color coded according to the key on the right, i.e. blue entries correspond to pairs of nodes having different time-series while yellow entries correspond to pairs with similar time-series.  Thus one can coarsely distinguish between high- and low-degree nodes.  This analysis on its own leads to a gross overestimate of the structural parameter, see panel (d). However, using this coarse classification and mathematical results on coupled systems (\ref{eq1}) satisfying assumptions (i), (ii) and (iii) makes it possible to find an approximation of the evolution rule at each node, see panel (e).  Using this additional information, one can obtain a  description of the network connectivity structure, see panel  (f). The network used in this figure has 1000 nodes, but only 50 are shown for clarity.}
\label{fig1} 
\end{figure}

{\bf Scale-free networks of coupled { bursting} neurons}.  {We consider coupled bursting neurons with excitatory synapses \cite{rulkov2001} in scale-free networks (see Methods for details on simulation of scale-free networks and formulation of the model). Our techniques work equally well for spiking dynamics. 

We generate a scale-free network with $N=10^4$ nodes  such that the probability of having a node of degree $k$ is proportional to $k^{-\gamma}$, where $\gamma=2.53$.  For this reconstruction we only need 2000 data points for each node.  Again, to  every pair of  time series $y_i$ and  $y_j$ we assign  a \emph{Pearson distance} $s_{ij}\ge0$ and the node intensity  $S_{i} = \sum_{j}s_{ij}$. The empirical distribution of the intensities $S_i$ approximates the degree distribution of the network, see Fig~\ref{fig1}(d). In the example here, the estimated structural exponent from the distribution of $S_i$ is $\gamma_{\rm est} = 3.1$, which yields a relative error of nearly 25\% with respect to the true value of $\gamma$ (see also  the plots in Figure \ref{fig2} a)). The functional network therefore overestimates $\gamma$, which has drastic consequences for the predicted character of the network. For example,  the number of connections of a hub for a scale-free network is  concentrated at $k_{\rm max} \sim N^{1/(\gamma - 1)}$, so the relative inaccuracy for the estimate $ k_{\rm est}$ of the maximal degree is $ k_{\rm max}/k_{\rm est}  = N^{1/\gamma - 1/  \gamma_{\rm est}}$, which is about  500\%. Such inaccuracy has important repercussions for the ability to predict the emergence of collective behaviour \cite{pereira2010,pereira2017}.  

The statistical measure used for the construction of functional network typically depend in a nonlinear way on the degrees, thus causing a distortion in the statistics. We will discuss the case of Pearson distance. Suppose that the signals $\{(y_i(t),y_i(t+1))\}$  are purely deterministic, $y_i(t+1)=g_i(y_i(t))$, and thus perfectly fall on the graphs of the functions $g_i$, that are determined by $\beta_i$ and so by the degree $k_i$ of the node. The Pearson distance $s_{ij}$ between the signal at $i$ and $j$ is going to be a number between 0 and 1, depending on how close these graphs are. Unfortunately this distance depends nonlinearly on the degrees $k_i$ and $k_j$. Devising another distance $s'_{ij}$  without the knowledge of the interaction, in general,  would still carry the nonlinear dependence on the degrees. Moreover,  once fluctuations from the network are included  as $y_i(t+1)=g_i(y_i(t))+\xi_i(t)$  the differences between time-series $\{(y_i(t),y_i(t+1))\}$ for different $i$s can be due to fluctuations rather than differences in the degrees.  So the decomposition of the rules in terms of interactions and fluctuations is essential to recover precise details of the real degree distribution.

\begin{figure}[h]
    \centering           
        \includegraphics[width=0.52\linewidth]{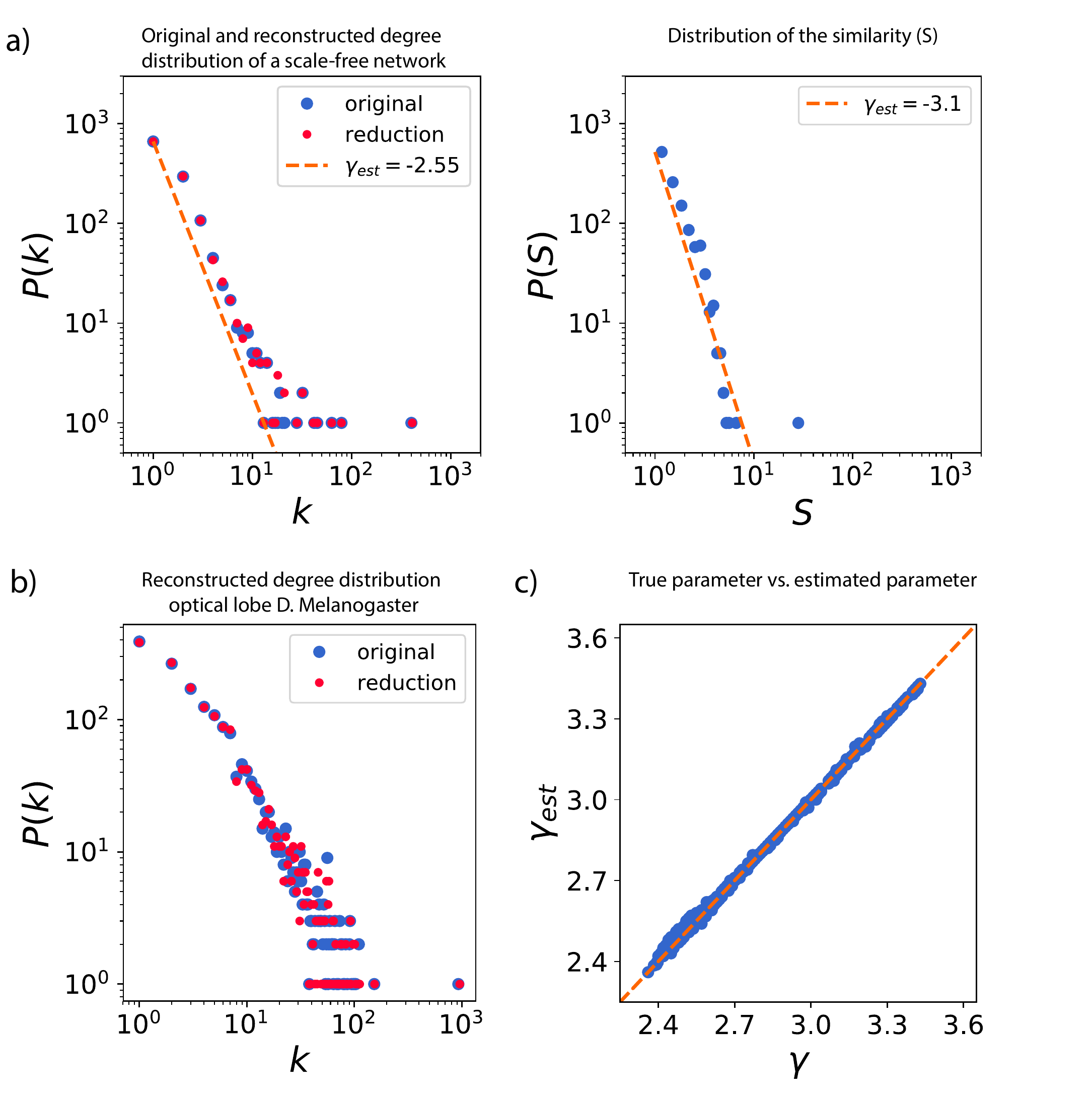}
\caption{ {\bf Reconstruction of  structural power-law exponents $\gamma$ of scale-free networks from data.}  We estimated $\gamma$ from the multivariate time series obtained by simulating chaotic coupled dynamics on random scale-free networks with degree distribution $P(k) \propto k^{-\gamma}$.  The three plots in panel (a) compare the functional and effective network approach in recovering the degree distribution. The first plot shows the real distribution of degrees, the second shows the distribution obtained with the functional network, and the last figure shows the distribution recovered with the effective network. The functional approach is able to recover that the degree distribution follows some power law and can distinguish low from high degree nodes. However, it does not recover the parameter of the power law accurately at all, see the second plot in panel (a). Much better estimates are obtained using the effective network, see the third plot in panel (a). Panel (b) shows the degree distribution of the original system (in blue) and that estimated from an effective model (in red) for the neural network in the optical lobe of Drosophila melanogaster. We obtained an accuracy of $3\%$ in the structural exponent of the tail. Panel (c) shows  the true exponent $\gamma$ versus $\gamma_{\rm {est}}$ obtained with an effective network from data for spiking neuron maps coupled with chemical synapses on networks with different values of $\gamma$. We generated 1000 networks, from  which the $\gamma_{\rm est}$ estimate is within $2\%$ accuracy. }
\label{fig2}
\end{figure}

The effective network provides a  better statistical description of the network structure. To compare with the functional network approach described above, we constructed an effective network of the same system tested for the functional network approach. The estimate for $\gamma$ from the effective network is $\gamma_{\rm est}=2.55$, which has an error of only $1\%$ ( Fig.~\ref{fig1}~(d)).  We repeat the analysis on a different network with different parameters $\gamma$ in the degree distribution. The estimated $ \gamma_{\rm est}$ values are shown in Fig.~\ref{fig2}~(c)  as a function of the true parameter $\gamma$. The relative error on the estimated exponent is within $2\%$.

{\it Real-world scale-free networks: the optic lobe of Drosophila melanogaster.} Effective networks can be used to provide estimates for real-world systems. As one example, we applied our method to data simulated from the neuronal network in the D. melanogaster optic lobe, which constitutes $>$50\% of the total brain volume and contains 1781 nodes \cite{takemura2013}. The degree distribution has a power-law tail \cite{garcia-perez2018}. We used spiking neurons with chemical coupling to simulate the multivariate time series, from which we constructed an effective model  capable of estimating the degree distribution with great accuracy (Fig.~\ref{fig2}~(b) and  Methods).

Additional applications of our procedure for different types of chaotic systems (doubling map, logistic maps, spiking neurons with electrical synapses, R\"ossler systems, and H\'enon maps) can be found in the Supplementary Materials, together with results demonstrating that  effective networks are robust against noise. 

\section*{Conclusions}
We have introduced an {\em effective network} obtained from time-series of a complex network (by observing the dynamics at each node). Our method complements the existing ones in two ways. Firstly, it encompasses the case of isolated chaotic dynamics. Secondly it deals with weak coupling among the nodes. Both cases are commonly found in applications. Key to the success of the reconstruction is the heterogeneity of the network which allows us to separate the contribution from the local dynamics and the interactions and thus perform a multi-level reduction. 

We have compared our procedure with methodologies  most relevant for the systems we considered. We have thus excluded results tailored to specific experimental setups or dynamics (binary dynamics coupled on networks \cite{li2017universal}, and see \cite{wang2016} for a review). We also did not consider methods that rely on measurements  obtained intervening on the system with controlled inputs (see e.g. \cite{nitzan2017revealing}) and restrict our attention to the case where the time-series is recorded under constant conditions. When the coupling is strong, sparse recovery can be applied \cite{brunton2015}. On the other hand, when the coupling is weak sparse recovery cannot distinguish small parameters from those that are identically zero thus misidentifying connections between nodes as it can be observed in Figure \ref{SR}. 
 Also model-free methods  are ill-suited to analyse weakly coupled chaotic dynamical systems studied here. Indeed, the influence of a single pairwise interaction on the time-series is so weak that its effect on the correlation of the measurements can hardly be detected by direct measurements.

Our reconstruction works when no collective dynamics is induced by the weak coupling.  When the rich-club synchronizes.
 A synchronous rich-club induces correlation between the fluctuations in the communities. This happens because two nodes in different communities but coupled to the rich-club would receive a similar forcing. Since we use the fluctuation to identify the community structure, the method would identify these nodes as belonging to the same community. 
%
While our method works well for various isolated dynamical systems, in two scenarios it didn't provide a good reconstruction. First, when the local dynamics is the Lorenz with classical parameters. In this case, the passage near a fixed points smears the fluctuations. Second, in the bursting dynamics, when the quiescent state is too long. This failure seems to be related as well to the long passage near a fixed point. Our method works with large networks, up to $~10^5-10^6$ nodes after which the computation time can be long. In fact, the number of operations to fit the the dynamics at all nodes, as in point (i), is proportional to
Number of nodes $\times$  (Length\,of\,the\,time\,series) $^\beta$,
where the power $\beta$ depends on the optimization algorithm adopted.
The number of operations to determine the correlations between the fluctuations, point (iii) above, is of the order of 
(Number\,of\,nodes) $^2$ $\times$ Length\,of\,the\,time\,series.
The effective network exploits the network heterogeneous structure.  This allows to single out different dynamical behaviours across the network and use this information to reconstruct the local dynamics and the effective interaction function. 
%
%
%
To obtain the community structures we use that  certain noise terms associated with the time series at two nodes in the same community are correlated. By collecting data when the network is far from critical transitions, our method can reconstruct the effective network and enables predictions regarding  critical transitions occurring in a network.

\newpage
\noindent {\LARGE \bf Methods}

\subsection*{Random scale-free networks}

A scale-free network has degree distribution $P(k) =  C k^{-\gamma}$, where $\gamma>0$ is the characteristic exponent and $C$ is a normalising constant.  We use a random network model $\mathcal{G}(\bm{w})$ which is an extension of the Erd\"os-R\'enyi  model for random graphs with a general degree distribution \cite{chung2002}. Given ${\bm w} = (w_1 , w_2 ,  \cdots, w_n )$, each potential edge between $i$ and $j$ is chosen with probability  $ p_{ij} = w_i w_j \rho, $  and where $\rho = (\sum_{i=1}^n w_i)^{-1}.$ To ensure that $p_{ij} \le 1$, we consider  $\max_i w_i  \rho \le 1.$ Then $w_i$ is the expected degree of the $i$th node.  Taking $w_i = ci^{-1/(\gamma-1)}$, where $\gamma \ge 2$ and $c = \frac{\gamma - 2}{\gamma - 1}dn^{1/\gamma - 1}$ with $d$ denoting the mean degree,  the distribution of expected degrees follows a power law, and so $P(w) \propto w^{-\gamma}$ is the expected exponent of power-law distribution \cite{chung2002,chung2003}.  If the random graph generated is not connected, we use only the largest connected component (the giant component). 

{\it Finding power-law distribution parameters from empirical data.}  For  estimation of the power-law distribution parameters, we use the maximum likelihood estimator first introduced by Muniruzzaman  \cite{muniruzzaman1957}, which is equivalent to the well-known Hill estimator~\cite{hill1975}. After that, we test the reliability between the data and the power law by using the goodness-of-fit method. If the resulting $p$-value is larger than 0.1, the power-law estimation is an appropriate hypothesis for the data. A complete procedure for the analysis of power-law data can be found in Ref.~\cite{clauset2009}.

\subsection*{Functional networks}
\label{sec:functional_networks}

For networks composed of chaotic oscillators, building the functional network from the 
standard Pearson correlation between time series gives no meaningful results because of the 
decay of correlation intrinsic to the chaotic dynamics. Hence, functional networks are built using a Pearson distance $s_{ij}\ge 0$,  which describes the proximity of the dynamics (i.e. the {\lq}attractor{\rq}) at two nodes $i$ and $j$.  To do this,  we consider the time series  ${z}_i(t):=({y}_i(t),{y}_i(t+1))$, $t=0,\dots,T-1$ reordered in $z_i^{\rm lex} (t)$ according to the lexicon order; that is, according to the magnitude of the first component of  $z_i(t)$. Then, let $r_{ij}$ be the Pearson correlation,  $r_{ij}=$ Cor$(z_i^{\rm lex}, z_j^{\rm lex})$, so  that $r_{ij}=1$ indicates that the attractors at nodes $i$ and $j$ fully agree.  
Finally, define the Pearson distance $s_{ij}=1-|r_{ij}|$ so that $s_{ij}=0$ indicates full agreement of the dynamics and $s_{ij}>0$ measures the difference between the attractors.

The intensity $S_i =  \sum_{j}s_{ij}$  approximates how many nodes have a dynamical rule different from $i$ and helps to distinguish between poorly connected nodes and hubs. Since  most of the network is composed of poorly connected nodes with similar evolution rules, they exhibit a smaller $S_i$ than high-degree nodes, which are scarcer and have different dynamics  from the low-degree nodes. In the case of a scale-free network  where the degree distribution is monotonic, the empirical distribution of $S_i$ is also expected to have the same trend. 

\subsection*{Effective network}

{\it Dimensional reduction over finite time scales}. To construct an effective network, we combine statistical estimation with mathematical results for high-dimensional dynamical systems that are based on theorems describing the evolution of dynamical systems coupled on a heterogeneous networks. 
 
The  assumptions we make are: (a)   the dynamics determined by $\bm F$ is ergodic, (b) the network is heterogeneous in the sense that most nodes are low degree and a few hubs make many connections, and (c) the coupling strength $\alpha$ is small so that $\alpha \max_i k_i = O(1)$. These assumptions  play a role in the reduction process. Indeed, (c) means that the contribution of the low-degree nodes to the dynamics of the hub nodes is an average, (b) means that the ergodic behaviour of low-degree nodes is similar to that of $\bo F$, and finally, (a) means that we obtain a hierarchy of reductions with  different dynamics (depending on connectivity) from which  the structure of the network can be inferred. 

In heterogeneous networks,  and up to a small error (relative to the network size), the dynamics at each node evolves according to a low-dimensional dynamical system for a certain time scale where the effect of the interaction of one node on the entire network is averaged to a give net interaction. This is rigorously proven in \cite{pereira2017} under some additional technical assumptions.
\def\kapppa{k}
One can write the interaction term in equation (\ref{eq1}) at node $i$ as 
\begin{eqnarray}\label{dimred2}
 \alpha\sum_{j=1}^{N} A_{ij} \bm{H}(\bm{x}_j(t),\bm{x}_i(t)) =  \alpha k_i \bm{V}(\bm{x}_i(t))+ \alpha k_i\bm{\xi}_i(t)
\end{eqnarray}
\noindent  where
\begin{eqnarray}
\bm V(\bm x)=\int \bm H(\bm x, \bm y)\, d\mu (\bm y)  \nonumber
\end{eqnarray}
is the averaged interaction in terms of the invariant measure $\mu$  for the isolated dynamics and $\beta_i = \alpha k_i$ where $k_i$ is the degree of node $i$.  The term  $\bm{\xi}_i(t)$ is determined by the states of the $i$th node neighbours. The low-dimensional system, called the {\it reduced system}, plus a fluctuation term thus reads as
\begin{eqnarray}\label{Eq:Red}
\bm{x}_i(t+1) = {\bo G_i}(\bo x_i(t)) +\alpha k_i \bm{\xi}_i(t)
\end{eqnarray}
\noindent
where 
${\bo G_i}(\bm{x})=\bm{F}_i (\bm{x}) + 
\alpha k_i \bm{V}(\bm{x}), 
$ for $i=1,\dots,N$. Using the chaotic and ergodic properties of the dynamics and concentration inequalities, we can  show that $|\alpha k_i\bm\xi_i(t)|$ is small over exponentially large time scales, in terms of the system size, and for most initial conditions. Not all structures of the graph will follow this reduced equation. If a fully connected graph is  weakly coupled to a  scale-free network, then it can develop  dynamics different from those of the reduced model.  In this case, we cannot use the physical measure $\mu$  in the expectation; however, the reduction is still valid whenever $\mu$ is substituted with a  measure that satisfies self-consistency relationships (see below).  

\section*{k-Fold Cross-Validation} 

\noindent Cross-validation is a resampling method used to evaluate dynamical models on a limited data sample. The parameter $k$ refers to the number of groups that a given data is to be divided into. We use $k=10$, since it is known that this value produces the lowest bias in several tests, so that the method can be called 10-fold cross-validation. The procedure is the following: First, randomly shuffle the data and divide it into k groups. Hold few groups to test the system, use the remaining ones as  training data. In our case, we used 1 group to test the system and 9 groups to train our model. We fit a model on the training set and evaluate it on the test one. According to the evaluation scores, we summarize the dynamical properties of the model. Further details on how  this standard procedure works can be found in  \cite{james2013}.

\section*{Reconstruction of the effective network representation from data}

{\bf Step 1 : Reduced dynamics.}  Given the multivariate time series $\{y_1(t), \dots, y_N(t) \}$, we consider each time series separately and then perform a Takens reconstruction of the attractor for each time series $y_i(t)$.  The reduction guarantees with high probability that the dynamics is low dimensional. If the time series is high dimensional, we discard it, and otherwise, once we are in the appropriate dimension, we continue to the next step.  Many cases can be captured by a one-dimensional map and Takens reconstruction yields the return map
\begin{eqnarray}
y_i(t+1) = g_i (y_i(t)) \nonumber
\end{eqnarray}
where $g_i : \mathbb{R} \rightarrow \mathbb{R}$. In fact, even when the local dynamics is  three dimensional, by introducing a Poincar\`e section, we can sometimes reduce the analysis to one dimension (see Supplementary Materials for details).

\noindent
{\bf Step 2: Isolated dynamics  and effective coupling function}. We first obtain a model for the isolated dynamics. Here we describe the procedure when the reduced dynamics can be modelled by a one-dimensional map, since the method works similarly in the higher-dimensional setting as shown in the Supplementary Materials.

To estimate the deterministic rule $g_i$ for each time series, we  use a $10$-fold cross-validation with $90\%$ of the time series for training and $10\%$ for testing (see k-Fold Cross-Validation {above}).  To obtain a model for the isolated dynamics we first build a functional network (see Functional Networks section in Methods). 
%
%
For the low-degree nodes, $\alpha \kapppa_i v$ is negligible and the dynamics at the low-degree nodes are  close to $f$. We identify these nodes by analysing the distribution of $S_i$. We use the top $N_{\rm top}$ nodes of the highest intensity nodes to obtain a proxy for the isolated dynamics. We then average the  rules obtained at those nodes to get $\langle g \rangle \approx f$. The choice of $N_{\rm top}$ is not fixed and depends upon the number of nodes and the fluctuation  $\sigma_g^2 = \langle (g_i - \langle g \rangle)^2 \rangle$, which is averaged from the $N_{top}$ highest intensity nodes. A good heuristic to choose $N_{\rm top}$ is $\sigma_g^2 / N^{1/2}_{\rm top} \ll 1$, as we have used here.

{\it The effective coupling function}. 
Since $\alpha \kapppa_i  v= g_i - \langle g \rangle$, analysing the family $\{ g_i - \langle g \rangle \}_{i=1}^{N}$  can yield the shape of $v$ up to a multiplicative constant
via a nonlinear regression by imposing that $g_i - \langle g \rangle$ and $g_j - \langle g \rangle$ are linearly dependent. The choice of the base function for the fitting is supervised and depends on the particular application (see Supplementary Material for additional examples). 

\noindent
{\bf Step 3. Network Structural Statistics}.
After selecting a $v$ that satisfactorily approximates $g_i- \langle g \rangle$ up to a multiplicative constant over all indices $i$, we estimate the parameter $\beta_i$  using a dynamic Bayesian inference. Because the fluctuations $\xi_i(t)$ are close to Gaussian we use a Gaussian likelihood function and a Gaussian prior for the distribution of the values of $\beta_i$, and hence obtain equations for the mean and variance. We split the data into epochs of $200$ points and update the mean and variance iteratively.

\noindent
{\bf Matching index:} Recall that the degree of node $i$ is $ k_{i} = \sum_{j}^N A_{ij}$ and count the number of neighbours it has.  We now define the \textit{matching index} of a graph~\cite{zamora-lopez2010}. Consider the neighbourhood of node $i$ as $\Gamma (i) = \{j \in \{ 1, \dots, N\} \, | \,  A_{ij} = 1\}$. This is the set of nodes that shares an edge with the node $i$. The matching index of nodes $i$ and $\ell$ is the cardinality of the overlap of their neighbourhoods  $\mu_{i \ell} = |\Gamma(i) \cap \Gamma(\ell)|$. We are interested in the normalised matching index:
\begin{eqnarray*}
\widehat{\mu}_{i\ell} = \frac{|\Gamma(i) \cap \Gamma(\ell)|}{|\Gamma(i) \cup \Gamma(\ell)|} 
\end{eqnarray*}
or equivalently
in terms of the adjacency matrix
\begin{eqnarray*}
\widehat{\mu}_{i \ell} 
= \frac{(A + A^2)_{i \ell}}{k_{i} + k_{\ell} - (A + A^2)_{i\ell}}.
\end{eqnarray*}
Clearly $\widehat{\mu}_{i\ell} = 1$ if and only if $i$ and $l$ are connected to exactly the same nodes, i.e., $\Gamma(i) = \Gamma(\ell)$; whereas $\widehat{\mu}_{i \ell} = 0$ if nodes $i$ and $\ell$ have no common neighbours. 

\noindent
{\bf Community structures.} Once we have obtained the rules $g_i$, we filter the deterministic part of the time series $y_i$ and access the fluctuations $\xi_i$. We decompose the fluctuations  $\xi_i = \xi_i^{\rm c} +  \xi_i^{\rm o}  $ where $\xi_i^{\rm c}$ is the fluctuation of the local mean field from nodes in the cluster containing  $i$, and  $\xi_i^{\rm o}$ is the contribution from outside the cluster. Since a node makes most of its connections within its cluster, $\xi_i^c\gg\xi_i^{o}$ with high probability,  and thus if $i$ and $j$ belong to the same cluster $ \mbox{Corr}(\xi_i , \xi_j)=\mbox{Corr}(\xi^c_i , \xi^c_j)$. Next, we notice that the common noise is generated by the common connections between nodes $i$ and $j$. In fact, for fixed isolated dynamics and coupling function 
\begin{eqnarray*}
\mbox{Corr}(\xi^c_i , \xi^c_j) \propto \widehat{\mu}_{ij}.
\end{eqnarray*}
It is well known that in the cat cerebral cortex nodes in the same community has a high matching index while nodes are distinct communities has a low matching index. In fact, this tends to be typically in modular networks \cite{zamora-lopez2010}. Therefore, for nodes in distinct clusters the component $\xi_i^{\rm c} \approx 0$, so $ \mbox{Corr}(\xi_i , \xi_j) \approx 0.$ Hence, we can recover the network structure by performing a {\em covariance analysis of the noise}.

\noindent
{\bf Filtering out the deterministic part}. 
Filtering out the deterministic part plays a major role in recovering community structures. 
Suppose we have two signals of the form $y_i(t) = {Y}_i(t) + \zeta(t), \quad i=1,2 $,  where ${Y}_i$ is independent  of $i$ and $\zeta(t)$ is a common noise term. 
$Y_i$ represents the superposition of the deterministic chaos and the independent fluctuations of the network setting. For the correlation, we have
\begin{eqnarray*}
\mbox{Corr}(y_i , y_j) \approx \frac{\mbox{Cov}(\zeta,\zeta)}{\sigma^2_{Y_1} \sigma_{Y_2}^2}
\end{eqnarray*}
\noindent
Hence, the large values of the variance of the time series ($\sigma_{y_i}\approx \sigma_{Y_i}\gg \sigma_{\zeta}$) suppress the contribution of the common noise, and an analysis solely based on the the original time series $y_i$ will overlook the common noise contribution. In fact, 
in the real system, we cannot assume ${Y}_i$ to be independently distributed. However, if the system is sufficiently chaotic, it will exhibit fast decay of correlations.

\section*{Generating the connectivity structure}
 To generate the adjacency matrix from a given degree distribution, we use the {\it configuration model} \cite{newman2003}. We know the number of stubs (half-links) for each node and the model randomly connects these stubs (half-links) to each other and generates an adjacency matrix that respects the given distribution.

\section*{Dynamical Models}

We tested the effective network on a wide range of dynamics and networks satisfying assumptions (a)-(c). We focused on scale free networks and on rich club networks to evaluate the performance of the reconstruction.  As a benchmark model, we restricted our attention to a neuron dynamics. Further models can be found in Supplementary Material.

\section*{Neuron Model}

We use  Rulkov maps to model spiking and bursting neurons \cite{rulkov2001}. In this setting, the variable $\bm{x}$ at each node is two dimensional and we will denote  $\bm{x} = (u, w)$. The local uncoupled dynamics is $\bm{F}(\bm{x}) = (F_1(u,w),F_2(u,w))$ with 
\begin{eqnarray*}
F_1(u,w) = \frac{\beta}{1+u^2} + w
\quad\mbox{ and }\quad
F_2(u,w) = w - \nu u - \sigma.
\end{eqnarray*}
The variable $u$ represents the membrane potential of a neuron, and, in this case, $u$ is the state variable measured by the observed time series $y_i(t)$; that is, $\phi(\bo x)=u$. Different combinations of parameters  $\sigma$  and $\beta$  give rise to different dynamical states of the neuron, such as resting, tonic spiking, and chaotic bursts.  To test our procedure we considered two cases where $\sigma=\nu=0.001$ and $\beta=5.9$, which correspond to tonic spiking and $\beta=4.4$ for bursting.

{\it  Chemical synapsis.} Synaptic coupling occurs only through the membrane potential $u$. That is, $ \bm{H}(\bm{x}_i, \bm{x}_j) = (h(u_i,u_j),0)$ with  $h(u_i,u_j) = (u_i - V_s) \Gamma (u_j)$, 
where 
\begin{eqnarray*}
\Gamma(u_j) = 1/ (1+\exp\{ \lambda( u_j - \Theta_s )\} )
\end{eqnarray*}
and $V_s$  is a parameter called reverse potential. Choosing $V_s > u_i(t)$, the synaptic connection is excitatory. We take $V_s = 20$, $\Theta_s = - 0.25$, and $\lambda=10$.
Applying the dimensional reduction in \eqref{dimred2}, we obtain a map with averaged value of the interactions
$
\bar F_1(u) = F_1(u) + \alpha k_i \langle \Gamma \rangle (u - V_s). 
$
One can \emph {a posteriori} verify that $ \langle \Gamma \rangle = \int \Gamma \mu(dx) $,  where $\mu$ is the physical measure for the map $F_1$ and can be estimated from the dynamics of the low-degree nodes. 

{\it Electrical synapsis.} Again, the synaptic coupling occurs only through the membrane potential $u$. That is, $ \bm{H}(\bm{x}_i, \bm{x}_j) = (h(u_i,u_j),0)$ with  
\begin{eqnarray*}
h(u_i,u_j) = u_j - u_i
\end{eqnarray*}
so the coupling depends only on the difference of states.

\subsection*{The cat cerebral cortex}

The dataset for connections in the cat cerebral cortex  consists of 53 geographic cortical areas investigated in detail with data-mining methods and was taken from \cite{scannell1993,scannell1995}. The network is  organised into four functional areas: visual, auditory, somatosensory-motor, and frontolimbic.

\noindent
{\bf Functional network.} Several methods can be used to create a functional network to recover the clusters. The approach is to compute pairwise correlations of all nodes and  to consider the similarities as interactions between nodes. This method is not appropriate for weakly coupled chaotic systems since they exhibit exponential decay of correlations.

\noindent 
{\bf Effective network representation.}  The connectivity matrix can be estimated using the correlation matrix $ \rho_{ij} = \mbox{Corr}(\xi_i , \xi_j)$.  The adjacency matrix $\bm{A}$ can be computed by thresholding the correlation matrix as $A_{ij}= \Theta(\rho_{ij} > \tau)$, where $\Theta$ is  the Heaviside step function and $\tau=0.5$ is an arbitrary threshold. Note that we tested different thresholds with values from 0.3 to 0.6 and obtained similar results.

\noindent 
 {\bf Finding communities.} Given the reconstructed connectivity matrix $\bm{A}$, we used the modularity-based Louvain method \cite{blondel2008} to detect communities. 
 
\noindent
{\bf Finding rich-club members.}  Colizza {\it et al.} developed an algorithm to detect members of the rich-club  \cite{colizza2006}. The algorithm provides a rich-club coefficient $\phi(k) \in [0,1]$  for each degree value  in the network. We considered  nodes with $\phi(k) > 0.8$ to be the members of the rich-club.

\section*{Predicting critical transitions}

Once we reconstruct the relevant information, we can perform numerical analysis of the recovered equations 
to explore the dynamics of the systems outside the observed range of parameters. Here we explain how to gather the information for a theoretical prediction of the critical transition.


{\bf Burst synchronization.} We introduce the slow variable as discussed in the main body of the manuscript.  To introduce a phase variable $\theta$ for the dynamics $y$ we first smooth the time series \cite{footnotesmoothing}. Then, we find the time $t_n$ of local maxima as the $n$th maximum point of the slow variable. We do this after performing a Gaussian filter to denoise slightly the data.  Finally, we introduce the phase variable $\theta$ as 
\begin{eqnarray*}
\theta(t) = 2\pi \left(\frac{t - t_n}{t_{n+1} - t_n} +t_n \right), \quad t_n < t < t_{n+1}
\end{eqnarray*}
as discussed in Refs. \cite{pereira2007}

\noindent
{\bf Reduction in the rich-club}. Nodes in the rich-club have degrees of approximately $\Delta$ and make $\kappa \Delta$ connections inside the rich-club and $(1-\kappa)\Delta$ connections to the rest of the  network. Following our reduction scheme, the interactions within and outside the rich-club  can be described by the expected value of the interactions with respect to the invariant measure associated with each of them.  Let $C$ denote the set of nodes in the rich-club, then the coupling term for the $i$th node in the rich-club
 is
\begin{eqnarray*}
\sum_{j} A_{ij} \bm{H}(\bm{x}_i,\bm{x}_j) =  \sum_{j \in C} A_{ij} \bm{H}(\bm{x}_i,\bm{x}_j) + \sum_{j \not \in C} A_{ij} \bm{H}(\bm{x}_i,\bm{x}_j)
\end{eqnarray*}
However,
\begin{eqnarray*}
\sum_{j \not \in C} A_{ij} \bm{H}(\bm{x}_i,\bm{x}_j) = (1 - \kappa)\Delta \int \bm{h}(\bm{x}_i, \bm{y}) d\mu(\bm{y}) + \bm{\xi}^o_i(t)
\end{eqnarray*}
where $\mu$ is the invariant measure for the nodes outside the rich-club. Hence,  for the rich-club we obtain 
\begin{eqnarray*}\label{RCeq}
\bm{x}_i(t+1) = \bm{q}_i(x_i(t)) +  \sum_{j \in C} A_{ij} \bm{H}(\bm{x}_i,\bm{x}_j) + \bm{\xi}^o_i(t),
\end{eqnarray*}
where $\bm{q}_i(\bm{x}_i(t)) = \bm{F}_i(\bm{x}_i(t)) + (1 - \kappa)\Delta\alpha \int \bm{H}(\bm{x}_i, \bm{y}) d\mu(\bm{y})$.

\noindent
{\bf Predicting the transition to collective behaviour }
Let us recall that when isolated 
$
u_i(t+1) = F_{1,i}(u_i(t))+ w_i(t)$ where $F_{1,i} \approx F_1$
and $ w_i (t+1) = w_i (t) + \mu (w_i(t) - 1)$. After some algebra we obtain
\begin{eqnarray}
\label{yeq}
w(t+1) = w_0 + \mu \sum_{n=0}^t (u(n) - 1)
\end{eqnarray}
Using the reduction Eq. (\ref{RCeq}), in the network we obtain
\begin{eqnarray*}
u_i(t+1) = F_{1,i}(u_i(t)) + u_i(t) + \alpha \Delta [ \langle u \rangle - u_i(t)] + \xi_i(t)
\end{eqnarray*}
where $i$ denotes the $i$th nodes in the rich-club, $\langle u \rangle$ is the mean membrane potential in the rich-club and 
$\xi_i$ are fluctuations. We fix two nodes  $y_i = u_i$ and $y_j = u_j$ in the rich-club. Let us introduce the disturbance
\begin{eqnarray*}
\zeta(t) = u_i(t) - u_j(t)
\end{eqnarray*}
Applying the mean value theorem and using that $F_{1,i} \approx F_1$ we obtain
\begin{eqnarray*}
\zeta(t+1) = DF_1 (x_i(t)) \zeta(t) + \mu \sum_{n=0}^t \zeta(n) - \alpha \Delta \zeta(t)
\end{eqnarray*}
and introducing a proxy for the dynamics of the slow variables  
\begin{eqnarray*}\label{ueq}
\eta(t) = \sum_{n=0}^t z(n)
\end{eqnarray*}
and considering 
\begin{eqnarray*}
\label{Approx}
\sum_{n=0}^t DF_1(x_i(n)) \zeta(n) \approx \lambda \sum_{n=0}^t \zeta(n)
\end{eqnarray*}
where we used that $\sum_{n=0}^t \zeta(n)$ is a slow variable.   We obtain
\begin{eqnarray*}\label{udyn}
\eta(t+1) = (\lambda - \alpha \Delta ) \eta(t) + \mu \sum_{n=0}^t \eta(n)  
\end{eqnarray*}
Recall that for the cat cerebral cortex $\Delta = 37$. We given the time series $\{y_i \}$ for $\alpha \Delta = 0.3$, we estimate $F_1$ using our method $i$ as the slow variables are constants over short time scales, and the obtain slow variables as a filter over the fast variables. Then, the parameters of the rich-club such as the mean number of connections among the rich-club in terms of the reduction and external connections in terms of strength of fluctuations (We will provide a throughly discussion about the estimation of these parameters for a fully chaotic system in the Methods).  From the data and using Eq. (\ref{Approx}) we estimate
$
\lambda = 1.42
$
and thus we obtain 
$
\alpha \Delta = 0.42.
$
At this critical value the slow variables tend the stay together due to the contraction in the dynamics. This is related to the onset of synchronization in the bursts, which is captured via a phase variable through the order parameter. 
}

\noindent
{\bf Sparse Recovery}. The method gives a way of recovering the equations from data by considering the evolution map as a linear combination of a well chosen basis of functions, called library. The main assumption is that many coefficients will be zero. That is, the vector of coefficients will be sparse \cite{brunton2015}.
Since we have $N$ nodes in our network, we consider the library 
$
\mathcal{L} = [ \phi_1(x_1),\phi_1(x_2),\cdots,\phi_1(x_N),\phi_2(x_1,x_N), \cdots , \phi_k(x_N,x_N) ]
$ as the set of basis functions. And denote the 
$
X = \left(x_1({t_{2}}), x_2({t_{2}}), \cdots, x_N(t_1), x_1(t_2),\cdots,x_N(t_n)\right)^*
$
where $^*$ denotes the transpose and we introduce
\begin{eqnarray*}
\Theta = 
\left(
\begin{array}{cccc}
    \phi_{1}(x_1(t_{1}))  & \phi_1(x_2({t_{1}})) &\cdots  & \phi_{k}(x_N({t_{1}}), x_N({t_{1}})) \\
 \phi_{1}(x_1(t_{2}))  & \phi_1(x_2({t_{2}})) &\cdots  & \phi_{k}(x_N({t_{2}}), x_N({t_{2}})) \\
     \vdots & \vdots & \vdots & \vdots \\
 \phi_{1}(x_1(t_{n-1}))  & \phi_1(x_2({t_{n-1}})) &\cdots  & \phi_{k}(x_N({t_{n-1}}), x_N({t_{n-1}})) \\
 \end{array}
\right)
\end{eqnarray*}

We then look for a solution of the system 
$
X = \Theta \Xi
$
where 
\[\Xi = 
\left(
  \zeta_{1,1} , \zeta_{1,2} , \cdots, \zeta_{1,N},  \zeta_{2,1} , \zeta_{2,2} , \cdots  \zeta_{kN}
\right)^*
\]is the vector of coefficients. The sparse recovery technique then solves the linear equation for $\Xi$ iteratively enforcing the sparsity of $\Xi$  by introducing 
$\sigma$ such that if $|\zeta_{ij}| \le \sigma$ we set such entry to zero \cite{brunton2015}. \looseness=-9

In our case, we will consider the scenario where the synapsis between neurons are electric and we generate the multivariate data for the cat cerebral cortex. We will assume we have knowledge of the coupling function so we can easily read the network structure from the sparse recovery. We choose a library of polynomials as $\phi_j(x_i) = x_i^j$ for $j=\{1,\cdots,5\}$ and the pairwise $\phi_j(x_i,x_k)$ to be homogeneous polynomials of degree two in the variables $x_i$ and $x_k$ for $i,k\in\{1,\cdots,N\}$. 

We solve the minimization problem and group dependencies of the network through the coefficient vector as the network structure \cite{brunton2015}. The strength of each connection is of order $\alpha \approx 0.015$. Hence we have chosen values of $\sigma$ close to this value. The reconstructed network does not identity the clusters correctly as can be seen by comparing the blue and red markers in  Figure \ref{SR}. Moreover, employing the technique of community detection shows that the probability of identifying a node in the correct cluster is around 45\%. This happens as each single connection is not strong enough. In fact, since the dynamics of the neurons are chaotic each connection plays a role of noise which leads to the misidentification.  Thus we have shown that the sparse recovery method is not appropriate in our setting.

\begin{figure}[h!]
    \centering           
        \includegraphics[width=1.0\linewidth]{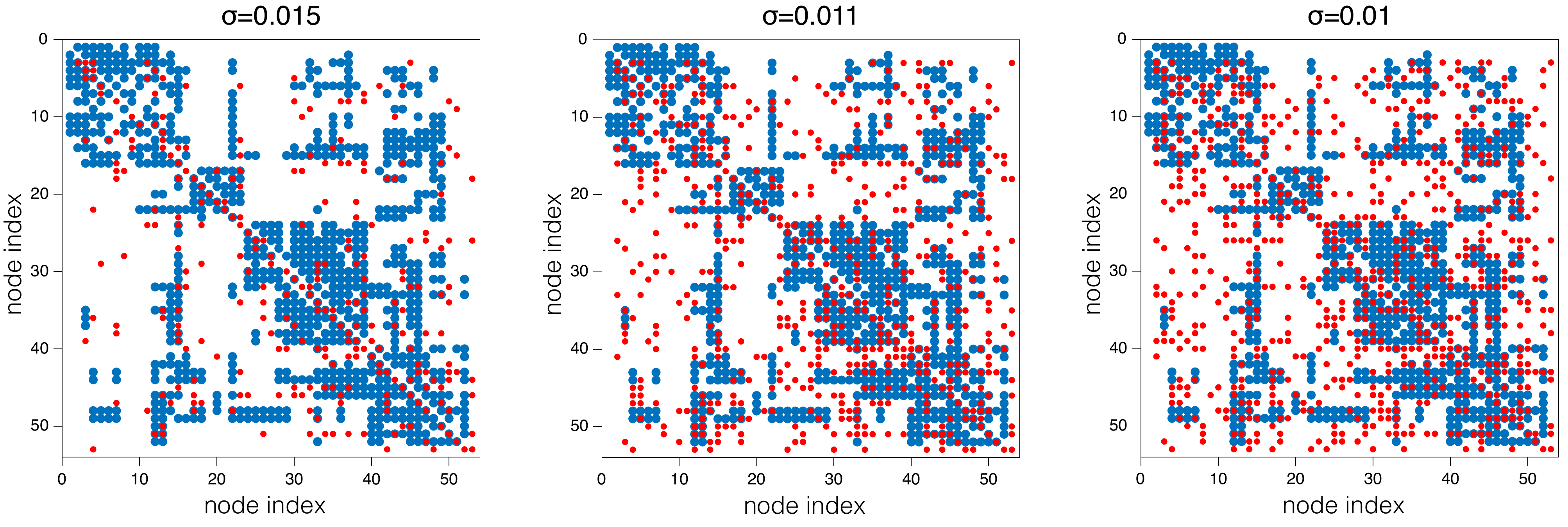}
\caption{Sparse recovery method is applied to the data  generated by bursting neurons electrically coupled on the cat cerebral cortex. Selecting the threshold parameter $\sigma$ in the method changes the reconstructed network. Here we show the results of sparse recovery method for different enforced sparsity $\sigma$. The nonzero entries of the original network's adjacency matrix are in blue.  The red filled circles represent the nonzero entries in the adjacency matrix of the network reconstructed with the sparse recovery method. As each connection is small in comparison with the isolated dynamics, the sparse recovery tends to neglect them. }
\label{SR}
\end{figure}

\textbf{Acknowledgements} We would like to thank Tomislav Stankovski, Chiranjit Mitra, Mauro Copelli, Dmitry Turaev and Jeroen Lamb for enlightening discussions. This work was supported in part by the Center for Research in Mathematics Applied to Industry (FAPESP Cemeai grant 2013/07375-0), the European Research Council (ERC AdG grant number 339523 RGDD), and the Serrapilheira Institute.

\textbf{Data Availability.} The connection matrices of cat cortex can be found at \url{https://sites.google.com/site/bctnet/datasets}. Connectivity matrix of Drosophila Medulla can be found at \url{https://neurodata.io/project/connectomes/}.

\newpage\noindent
{\LARGE \bf APPENDIX: Supplementary Material}

\section{Dimensional Reduction in Heterogeneous Networks}

Using the notation in the main body of the paper, we  present an informal statement of the theoretical result that suggests the reconstruction procedure. For a precise statement and details of the general setting see \cite{Tanzi}. The theorem has three main assumptions concerning the uncoupled dynamics $\bo F$, the network encoded in its adjacency matrix $A_{ij}$, and the reduced dynamics $\bo G_j$. We discuss a particular case and use it to illustrate the reconstruction step-by-step.

\begin{itemize}
\item[1)] {\bf The local dynamics must increase the distance between points} by a constant factor. A paradigmatic one-dimensional example is $ \bo F( x)=2 x$ mod $1$ that presents some of the most interesting characteristics of chaotic maps.
\item[2)]{\bf The networks are heterogeneous}, having nodes with disparate degrees. This assumption is made precise in \cite{Tanzi} by a set of conditions involving the size of the network $N$, the number of hub nodes $M$, the maximum degree of a hub node $\Delta$, and the maximum degree of a low degree node $\delta$. One instance in which these conditions are satisfied and that includes many situations of interest is the following. Suppose $M$ is constant or slowly increasing with the size of the network, $M\sim \log N$, $\Delta$ grows faster than the square root of the system size, $\Delta\sim N^{\frac{1}{2}+\epsilon}$, and $\delta$ increases slowly, but polynomially in the system size, $\delta\sim N^{\frac{\epsilon}{2}}$. Then for a sufficiently large $N$, the network satisfies the conditions.
\item[3)] {\bf The reduced dynamics must be hyperbolic}. This means that we assume the maps $\bo G_j$ to be either expanding (possibly by a non-constant factor), or to have a finite number of attracting periodic orbits. Every map can be perturbed by an arbitrarily small amount to obtain such an hyperbolic map \cite{Strien}.
\end{itemize}
Under these assumptions, we can prove the following approximation result

\begin{theorem}[\cite{Tanzi}]
For every hub node $j$, the dynamics at the hub is given by
\[
\bo x_j(t+1)=\bo G_j(\bo x_j(t))+\bo \xi_j(t)
\]
where $|\bo\xi_j(t)|<\xi$ for time $T$ with $1\leq T\leq \exp[C\xi^2\Delta]$, and a set of initial condition of measure $1-T/\exp[C\xi^2\Delta].$, where $C$ is constant in $\Delta$, and $\xi$.
\end{theorem}

Notice that one can pick the time scale $T$ exponentially large, but such that $T/\exp[C\xi^2\Delta]$ is very small so that, for large $\Delta$, the approximation result holds for very long time and for a large set of initial conditions.  The reduced dynamics given by $\bo G_j$  depends on the hub's state only. The term $\bo \xi(t)$ gives the fluctuations of the sum of all interactions from their expectation with respect to the physical measure of the unperturbed dynamics.


\section{Reconstruction of degree distributions}

{\bf Scale-free networks.} We create ensembles of scale-free network of $N=6000$ nodes with distinct structural exponents $\gamma$ following the same technique as in the main body of the manuscript. As in the main body, we only consider the largest connected component of the network (giant component). For each network realisation we compute the largest degree $\Delta$ and, for simplicity, we fix the coupling strength $\alpha / \Delta = 0.5$ throughout the rest of the section.  

\subsection{Doubling map}

Let us now apply our approach when $\bm{F}_i$ is a perturbed version of the doubling map. Since the dynamics is one dimensional, we denote $\bm{x} = x$ and $\bm{F}_i(\bm{x}) = f_i(x)$ with  
\[
f_i(x) = 2x + \varepsilon_i \sin 2\pi x \mod 1
\] and where we take $\varepsilon_i$ to be i.i.d. random variables (i.e. independent over $i$) uniformly distributed on $[0,10^{-3}]$. Likewise we write $\bm{H} = h$ with 
\begin{equation}\label{Eq:DiffCoup}
h(x_j,x_i) = \sin2\pi x_j-\sin2\pi x_i.
\end{equation}
We fixed $\alpha = 10^{-2}.$ Since the  unique absolutely continuous invariant measure for the doubling map is the Lebesgue measure $m$,  we have 
\[
\bo V(\bo x)=v(x) = \int h(y,x)d m(y)
\] 
yielding $v(x) = -\sin 2\pi x$ and the  reduced dynamics takes the form 
\begin{eqnarray}
x_i(t+1) = f_i(x(t)) - \alpha k_i v(x(t)) + \alpha \xi_i(t).
\label{eq:hubs2}
\end{eqnarray}
We aim to recover  the reduced dynamics and in particular $f$ and $v$ from the data  of a multivariate time series with $T=2000$ time steps. 

\begin{figure}[h]
      \centering
        \includegraphics[width=0.37\linewidth]{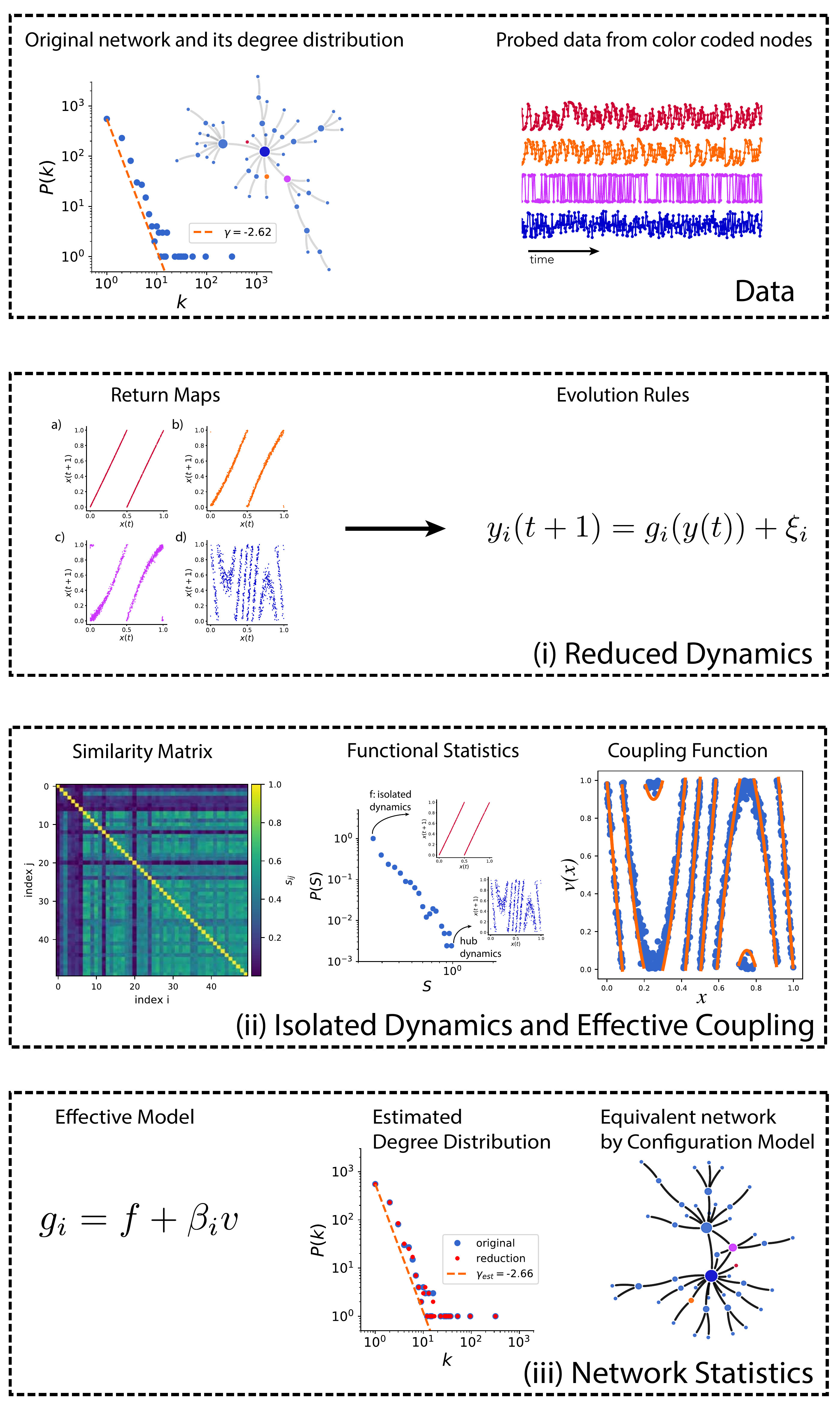}%
\caption{[Color Online] Step-by-step construction of an effective network for the doubling map. Our approach recovers the local dynamics of the doubling map, interaction function and statistical structure from the time series. Starting from datasets we uncover the approximate evolution rule for the time series using machine learning techniques. Such rules will be different for different nodes (depending on their degree) as shown in panel (i). Analysing the differences between these rules by means of a similarity analysis, we are able to obtain model of the isolated dynamics and an estimate the coupling function, see panel (ii).  Finally, by using the theory of dynamical systems and dimensional reduction, we are able to estimate the number of input each node receives and the community structures. We then create a random presentation of the network structure by using a network model such as the configuration model.}
\label{fig:Approach} 
\end{figure}

\noindent{\bf From data to Model.} We assume not to have access to the network structure and only measure the time-series $\{x_i(t)\}$ at each node, as illustrated in Figure \ref{fig:Approach} Data. By performing the steps described in the main body of the manuscript, we can recover an effective network. We now illustrate in detail our procedure in the case above of the doubling map.

\noindent
{\bf (i) Reduced dynamics.} From the time series observed at each node, we construct the attractor for that particular node. Computing the embedding dimension we notice that the dynamic at each site (node) is well described by a one dimensional map, $g_i(x)= f_i(x) - \alpha k_i v(x)$, up to a fluctuation $\xi_i$. Hence, to reconstruct the attractor it suffices to obtain the return map. Return maps at different nodes are shown in Figure  \ref{fig:Approach} (i).

\noindent
{\bf (ii) Isolated dynamics and effective coupling.} We start by introducing a similarity measure.

{\it Similarity of the time series.} Two time series  are considered almost the same, if whenever $y_i(t)$ and $y_j(t')$ are close then  $y_i(t+1)$ and $y_j(t'+1)$ are also close. This means that one considers the new times series  ${z}_i(t):=({y}_i(t),{y}_i(t+1))$, $z_{j}(t):=({y}_j(t),{y}_j(t+1))$, $t=0,\dots,T-1$, reshuffle them according to the lexicographical ordering (i.e. according to the magnitude of the first coordinate), and then take $s_{ij}$ to be the Pearson correlation distance between these reshuffled sequences. So if the time series at $i$ and $j$ are just out of phase, then, thanks to the reordering, their distance $s_{ij}=0$. We then calculate the intensities $S_i = \sum_{j} s_{ij}$ and use this to classify the nodes. The similarity matrix is shown in Figure \ref{fig:Approach}(ii).

Using the correlation distance described above, we determine for which nodes $S_i$ is minimal and in this way we identify the low degree nodes.  For the low degree nodes, the sequence $\{(y_i(t),y_i(t+1))\}_{t=0,\dots,T-1}$  lies close to the graph of the return map for $f$, and thus we obtain a good estimate for the isolated dynamics $f$, see Figure \ref{fig:Approach}(ii). Next, we apply the reduction to estimate the coupling function. 

To obtain the effective coupling $v$ from data, take a hub $j$ and consider the sequence $\{(y_j(t),y_j(t+1))\}_{t=1}^{2000}$. The resulting time series approximates the graph of $g_i$ and subtracting $f$ gives an approximation of the function $v$ up to a multiplicative constant (shown in Figure  \ref{fig:Approach}(ii)).

\noindent
{\bf (iii) Network Structural Statistics.} There are two ways to obtain information  about the statistics of the degrees $k_i$. The first one uses the noise variance. In fact, the size of the fluctuations $\xi_i$ depends on the number of connections the node $i$ makes, and with good approximation Var$(\xi_i) \propto k_i$. The second one uses  $f$ and $v$ recovered at the previous step. For every node $i$, choose $\beta_i$ to fit the time series $\{(y_i(t),y_i(t+1))\}_{t=0,\dots,T-1}$ with the map  
\[
g_i(y) = f(y) - \beta_i v(y).
\] 
The value of $\beta_i$ is obtained by Bayesian inference on the return maps is shown in the first panel of Figure \ref{fig:Approach}(i) as $f$ and $v$. The  distribution of  $\beta_i$ will be linear proportional to the degree distribution and so we can use it to obtain the the structural parameter. At this point we can construct an effective network with degree distribution close to that of the real network can be constructed with the configuration model as discussed in the main body of the manuscript. We can also check whether there are communities in the network by analyzing the covariance of the fluctuations $\xi_i$. 

In Figure \ref{gamma_est} a) we reconstruct from the distribution of $\beta_i$ the structural exponent $\gamma$ for 1000 scale-free network with exponents ranging from $2.4$ to $3.6$

\subsubsection{Robustness of the reconstruction under noise} 

\begin{figure}[h]
    \centering
        \includegraphics[width=0.8\linewidth]{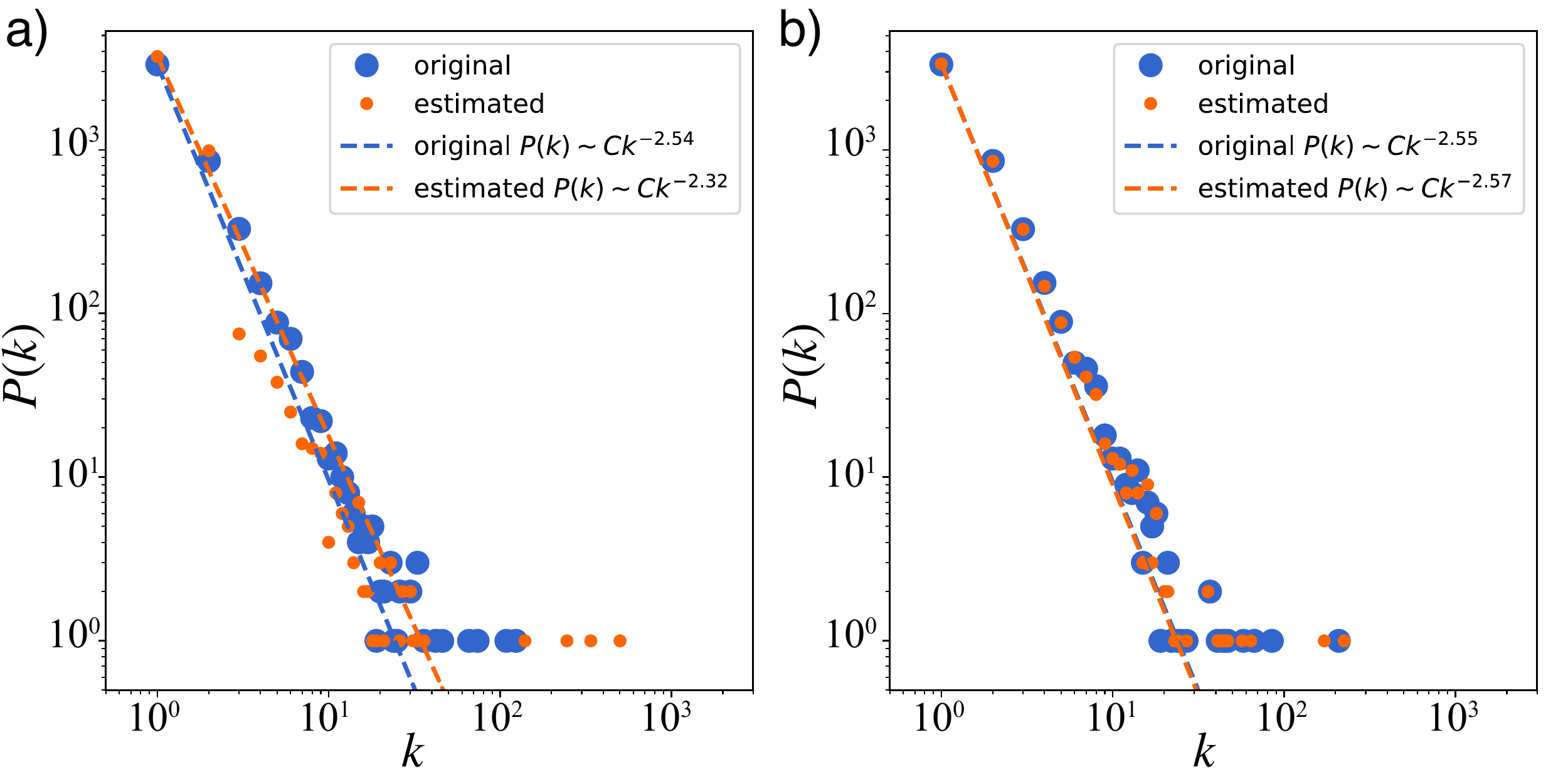}
    \caption{[Color Online] Effective networks are robust under random perturbations. When stochastic perturbations are moderate effective networks provide sharp estimates on the network structure. If the noise is large, the difference between the time-series at low and high degree nodes becomes blurred. In this case even the effective network underestimates the structural parameter $\gamma$. Inset a) shows the reconstruction of the degree distribution for $\epsilon_0=0.1$. Inset b) shows that the reconstruction is unaffected by the noise for $\varepsilon_0 = 10^{-3}$.}
\label{Doubling_noise}
\end{figure}

Adding some small independent noise  to the dynamics does not influence much the reconstruction procedure for the doubling map. This is a consequence  of stochastic stability of the local dynamics together with the persistence of the reduction \cite{Tanzi}. On the other hand, when the fluctuations become too large the reconstruction will underestimate the network structure. 
To illustrate these effects, we consider the randomly perturbed doubling maps
\[
x_i(t+1) = f_i(x(t)) + \eta_i(t) 
\]
where the random variables $\eta_i(t)$ are independent over $i$ and $t$, and identically distributed uniformly in the interval $[-\eta_0,\eta_0]$. Intuitively, as long as  $\eta_0 < \alpha \min k_i$  the reconstruction will go through as the noise fluctuation will not compete with the coupling term. Notice that we normalize $\| v \| =1$. This is illustrated in Figure \ref{Doubling_noise}. In inset a) the noise has a large support $\varepsilon = 0.1$, and in particular  larger than the coupling  $\alpha \min k_i = 10^{-2}$. As a consequence, the reconstruction understimate the number of low degree nodes. In inset b) the noise has a small support $\varepsilon = 10^{-3}$.

\subsection{Logistic Maps on Scale-Free Networks} 

\begin{figure}[h]
    \centering
    \includegraphics[width=0.6\linewidth]{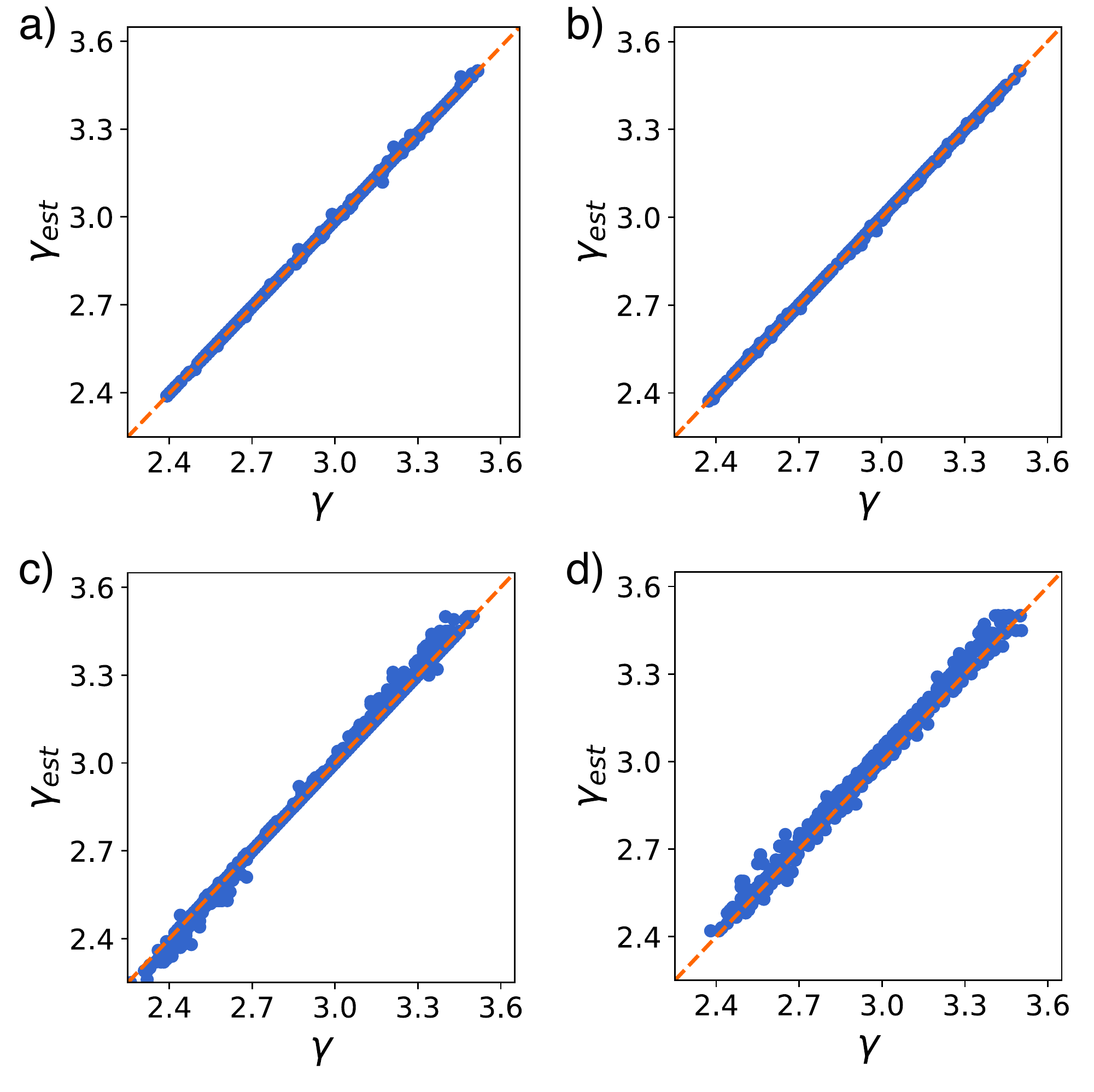}
%
    \caption{Recovering the structural parameter of scale free networks using effective networks. The power-law exponent $\gamma$ versus the estimated  $\gamma_{\text{est}}$ from data for 1000 distinct networks. Inset a) shows the reconstruction for coupled doubling maps with diffusive coupling. Inset b) shows the reconstruction for coupled logistic maps with Kuramoto interactions. Inset c) Spiking neurons with electrical coupling, and finally, inset d) shows the reconstruction for the H\'enon maps with the $y-$component diffusive coupled at the $x-$component. }
\label{gamma_est}
\end{figure}

Consider $M=[0,1]$, $\bm{x} = x$ and $\bm{F}_i(\bm{x}) = f_i(x)$ where  
\[
f(x) := 4 x(1-x),
\]
It is known that for such parameter, the map $f$ has an absolutely continuous invariant probability measure.  In contrast to the doubling map, a small perturbation of the logistic map might result in a map that has a non-smooth physical measure, namely the measure is a distribution given by the sum of Dirac delta masses at a finite number of points that constitute the periodic orbits of the system. This compromises the validity of the rigorous result analysis that we used in the case of the doubling map, for which small perturbations produced small changes in the invariant measure \cite{Strien}.  Nonetheless, as we will show, the reconstruction analysis gives good results in this case as well. We believe that this happens because our reconstruction only needs a finite number of points and gives a representation of the dynamics over finite time-scales. As a coupling function, we consider 
\[
h(x_j,x_i) = \sin(2\pi x_j-2\pi x_i)
\] 
and the reduction is given by $ x_i(t+1) = f(x_i(t)) + \alpha k_i v(x_i(t)),$
where $v$ is defined as above. Again, we consider the reduced model $g_i(x)=f(x) + \beta_i v(x)$ in terms of the free parameter $\beta_i = \alpha k_i$. Analysing the distribution of $\beta_i$ we access the distribution of the degrees. Results are presented in Figure\ref{gamma_est}b). 

\subsection{Spiking Neurons with Electrical Synapses}

We use the same models for the local dynamics of the spiking neurons which are described in the section Methods. We consider here a different coupling function. Denoting $\bm{x} = (u,w)$, the coupling is  
\[
\bm{H}(\bm{x_i},\bm{x_j})=\bm{E} ( \bm{x}_j - \bm{x}_i  ) = (u_j - u_i,0)
\]
Results of the reconstruction are presented in Figure \ref{gamma_est}c).

\subsection{Henon Maps on Scale-Free Networks} 

Using the same notation, $\bm{x} = (u,w)$, the coupled H\'enon maps are given by
 \[
    \bm{F}(\bm{x})=\left\{
                \begin{array}{ll}
                  1-1.4u^2 + w\\ 
                  0.3 w
                \end{array}
              \right.   \mbox{~ ~ and ~ ~ }
\bm{H}(\bm{x_i},\bm{x_j}) = 
\left\{
                \begin{array}{ll}
                  w_j-w_i \\ 
                  0
                \end{array}
              \right.
\]

We observe the dynamics of the first component  of the H\'enon map, that is, $ y = \phi(\bm{x}) = u$. 
In this case, the reconstruction starts by determining the dimension  
of the reduced system. Takens embedding reveals that the dimension is two for large time excursions.   Hence, we aim at learning a function of the kind 
 \[
 y_i(t+1) = g_i(y_i(t),y_i(t-1)) + \xi_i(t).
 \]
We use polynomial functions for the fitting and perform a 10-fold cross-validation.  
Just our theory implies that 
\[
g_i(y_i(t),y_i(t-1))  = f (y_i(t),y_i(t-1)) + \alpha k_i v(y_i(t),y_i(t-1)) 
\]
where $f$ models the isolated dynamics and $h$ the coupling. Again we obtain $f$ from the low-degree nodes via a similarity analysis. We learn the function $h$ by 
\[
\alpha k_i v(y_i(t),y_i(t-1))  = g_i(y_i(t),y_i(t-1))  - f (y_i(t),y_i(t-1))
\]
In our case, $v(y_i(t),y_i(t-1)) \propto y_i(t-1)$
%
%
and we can obtain the degree distributions, Fig.~\ref{gamma_est}d).

\subsection{Coupled Roessler Oscillators} 

Assume that the local dynamics is modelled by a R\"ossler oscillator \cite{Roessler}. The dynamics is now in continuous time and our method can also be applied by using a suitable Poincar\'e section. This gives an induced  map that describes the dynamics of the system at specific instants of time (when the system hits a selected subset of phase space). Denoting $\bm{x} = (x, y, z)^*$ the vector field is given by 
$
\bm{F}(\bm{x}) = (y - z, x + 0.2 y , 0.2 + z(x - 9) )
$ and the coupling function, assumed to be diffusive, is given by $\bm{H}(\bm{x}_i,\bm{x}_j) = \bm{E}(\bm{x}_j-\bm{x}_i)$, where $\bm{E}$  projects to the first component, i.e.,  $\bm{E}(x,y,z) = (x,0,0)$. So, our main equation reads as 
\[
\dot{\bm{x}} = \bm{F}(\bm{x}) + \alpha \sum_{j=1}^N A_{ij} \bm{E}(\bm{x}_j - \bm{x}_i)
\]
 We perform a numerical integration of the equations on the Rich-Club network using a 4th order Runge-Kutta with integration step $10^{-4}$ and get the data $\{ \bm{x}_i(t) \}_{t\ge 0}$.  Using a statistical analysis of the time-series of the state variables, we are not able to reveal the connectivity structure. 

\begin{figure}[h]
      \centering
        \includegraphics[width=1\linewidth]{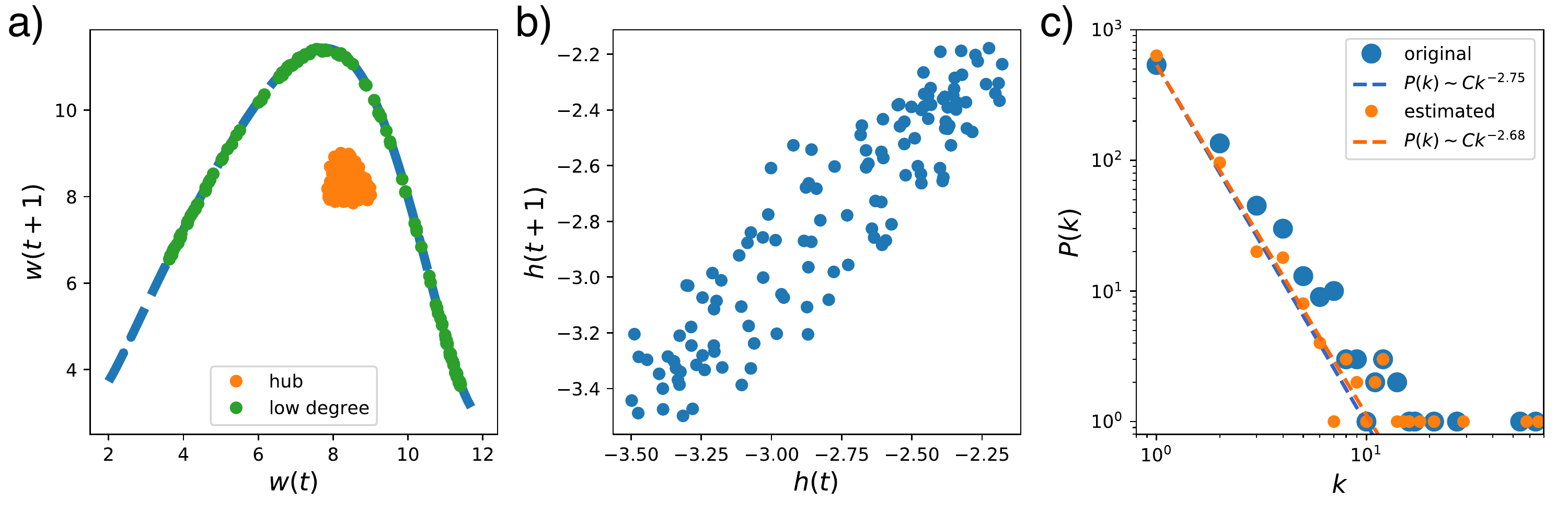}
\caption{Main estimations of the reconstruction for the R\"ossler systems. In inset a) we show the return maps obtained from the time-series of a hub and a low-degree node. Inset b) shows a return plot for the coupling function which is used to estimate the reduced dynamics and the degrees. Inset (c) shows  the power-law distribution of the degrees estimated from data and for the original network.}
\label{fig:Approach2} 
\end{figure}

The data is phase coherent, that is, taking a Hilbert Transform we can decompose the time series in terms of amplitude and phase we conclude that the spread in the phase variable is small and thus the return time to a given section is nearly constant.  So, we consider the Poincar\'e section defined by the maxima $w_i$ of the time series $x_i (t)$. This gives us a time series $\{ w_i(n)\} $ indexing all maxima. We then apply all the steps of the reconstruction procedure to this time series. Because of the coherent dynamics of the phase, the 
coupling form is preserved in the Poincar\'e section. The results of the network structure estimation are presented in Figure \ref{fig:Approach2}.

\section{Reconstruction of Rich-Club Mofits}


We report the performance of the method in the setting of a network of $100$ nodes having five clusters of $20$ nodes each. Four of these clusters are modelled as Erd\"os-Renyi random graph with  connection probability $p=0.3$. The remaining cluster is the integrating clusters with connection probability $p=0.8$.  The network resembles Fig. \ref{RC}.

\begin{figure}[t]
    \centering
        \includegraphics[width=6cm]{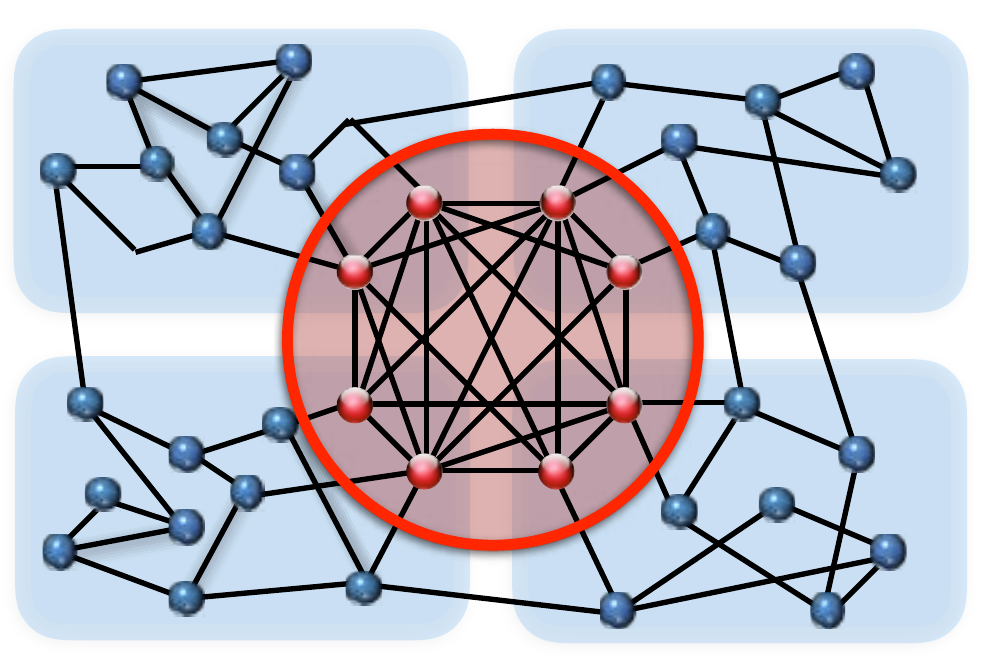}
        \caption{[Color Online] Illustration of the rich-club motif. The network is composed of communities and certain nodes from each community are highly connected among themselves forming an integrating cluster.}
\label{RC}
\end{figure}

\noindent
{\bf Filtering out the deterministic chaos.} We need to filter the contributions of the deterministic parts to reconstruct  the community structure. Indeed, for two nodes $i$ and $j$ in the same cluster, the signals have the form
$
x_i(t) = \tilde{X}_i(t) + \zeta(t), 
$
and $x_j(t)=\tilde{X}_j(t)+\zeta(t)$ where $\tilde{X}_i$ and $\tilde{X}_j$ is a superposition of the deterministic chaos depending on the variable at the node and  independent fluctuations coming from the rest of the network, while  $\zeta$ is common noise.  $\tilde{X}_i$ and $\tilde{X}_j$ have fast decay of correlations, depends on different sets of variables,  and for the sake of the following argument, can be assumed to be independent between each other and with $\zeta$. Under these assumptions, 
$ \mbox{Corr}( \tilde X_i(t) + \zeta(t) , \tilde X_j(t) + \zeta(t))  = \mbox{Var}(\zeta)/ \sigma_{x_i} \sigma_{x_j}$. Hence, the large values of the variance of the time series leads to strong suppression of the correlation coming from the small common noise $\zeta$.

\subsection{Condition for recovering  the community} 

In  general, the coupling function is a sum of terms 
$
h(x,y) = u(x) v(y).
$
 This leads to noise terms 
\[
\xi_i (t) = u(x_i) \left( \frac{1}{\Delta}\sum_{j} A_{ij} v(y_j) - k_i \int v(y) d\mu(y) \right)
\]
where $\mu$ is the physical measure of the local dynamics. Given $i$ and $j$ the sum can be split into connections commons the $i$ and $j$ and to the independent connections. So, for such $i$ and $j$ we can write 
\[
\xi_i  = u(x_i)[\zeta_i(t) + w(t)]  \mbox{~ and ~} \xi_j = u(x_j)[\zeta_j(t) + w(t)]
\]
where $w$ is the noise due to the common connections. Notice that $w$ has zero mean.  Let's estimate the covariance of the 
component $\xi_i$ and $\xi_j$. By abuse of notation we will omit the time index $t$ and write the covariance and by the previous computations 
$
\mbox{Cov}(\xi_i, \xi_i) = \mathbb{E}  [(u(x_i) w)( u(x_j)  w) ].
$
 After some manipulations, we obtain 
\begin{eqnarray}
\mbox{Cov}(\xi_i, \xi_i) &= & \langle u \rangle^2 \mbox{Var} (w) 
\end{eqnarray}
so, if 
$
\int u (x) d\mu(x) = 0, 
$ the correlation between the noise will vanish even though they have a common term. Thus, the 
generic condition in order for the above scheme to be able to recover communities is that $ \langle v \rangle \not=0$. If this condition is not met, the network reconstruction via the $g_i$'s is also not possible. We recall that $\langle v \rangle=0$ is not a generic condition and this is why the effective network approach works in most cases.

\subsection{Doubling Map on Rich Club Network}
\begin{figure}[h]
    \centering
        \centering
        \includegraphics[width=0.8\linewidth]{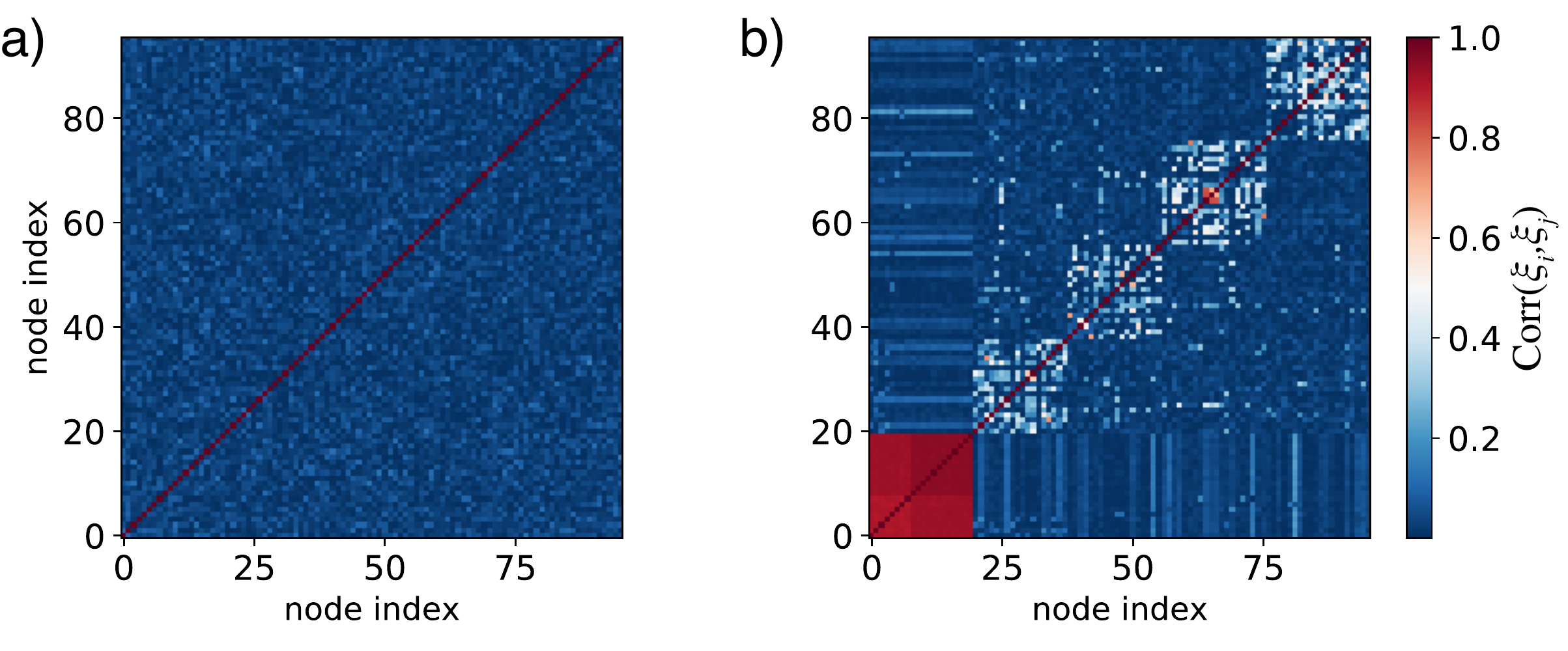}
    \caption{[Color Online] {Reconstruction of a rich-club motif for coupled doubling maps with diffusive coupling $h(x,y) = \sin 2\pi y - \sin 2\pi x$.} The left panel is a color map of the pairwise Pearson-correlation between the time-series $x_i$ and $x_j$. Due to the strong chaotic behaviour this analysis does not give any information on the community structures. In the right panel, we show the colormap of the noise correlation. The clusters are manifested in the correlation structure of the noise, namely, the four communities and the integrating cluster. }
\label{doubling}
\end{figure}

Using the doubling map with the diffusive coupling described in the above section on the reconstruction for scale-free networks, we test the community detection from time-series using the correlation between time series and correlations between the noise $\xi_i$. Then we apply our technique to recover the network structure. We show the results in Fig. \ref{doubling}

%
%

\subsection{Logistic Maps on a Rich Club Network}
\begin{figure}[h]
    \centering
        \centering
        \includegraphics[width=0.8\linewidth]{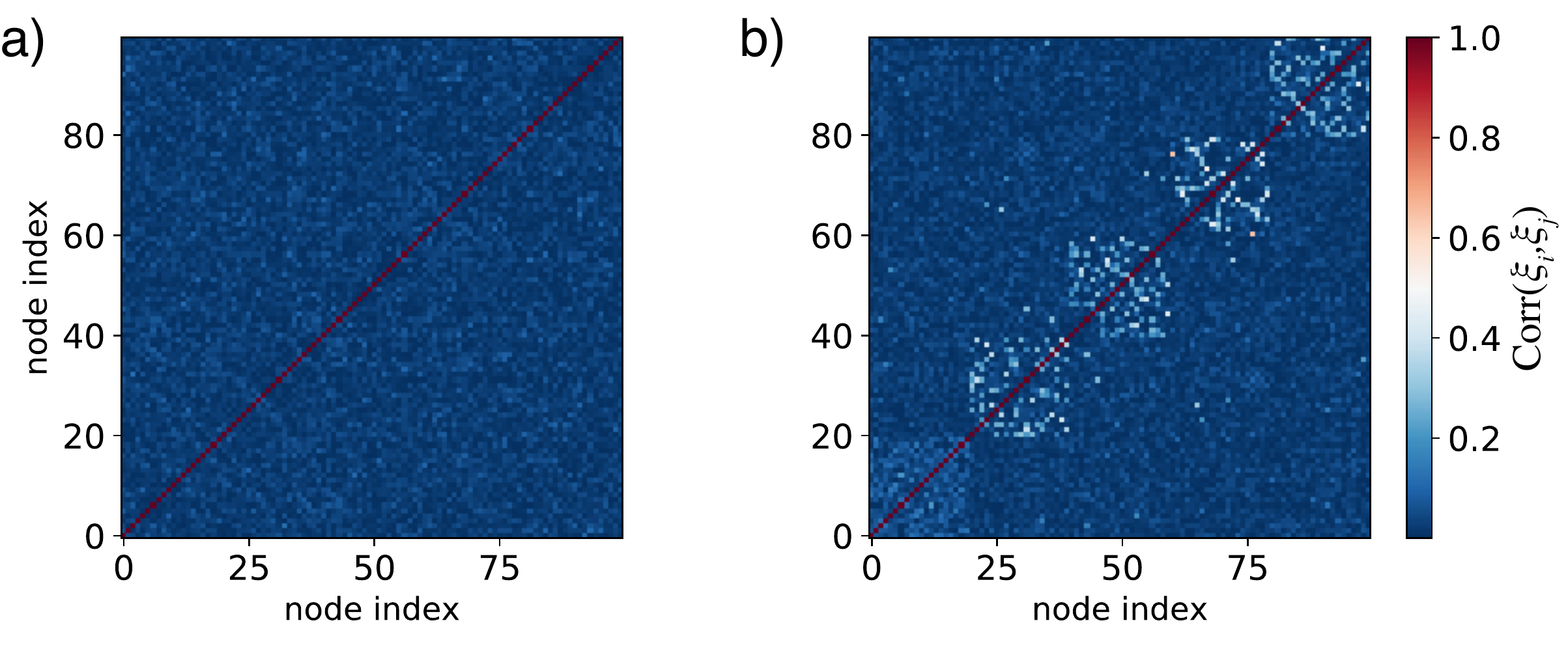}
    \caption{[Color Online] Reconstruction of rich club for logistic maps with Kuramoto coupling $h(x,y) = \sin 2\pi( y - x)$. The left panel shows a color map of the pairwise Pearson-correlation between the time-series $x_i$ and $x_j$. In the right panel, we show the colormap of the noise correlation that reveals the community structure and, in particular, the integrating cluster.}
\label{logistic}
\end{figure}

The dynamic of the logistic map and the coupling are described in the above section on the reconstruction for scale-free networks. We test the community detection approach on time-series using the correlation between time series and correlations between the noise. To this end, we fix the coupling strength $\alpha=10^{-4}$ and simulate the network dynamics. Then we apply our technique to recover the network structure. We show the results in Fig. \ref{logistic}.

\subsection{Spiking neurons coupled with Electric Synapsis  on a Rich-Club} 
\begin{figure}[h]
    \centering
        \centering
        \includegraphics[width=0.8\linewidth]{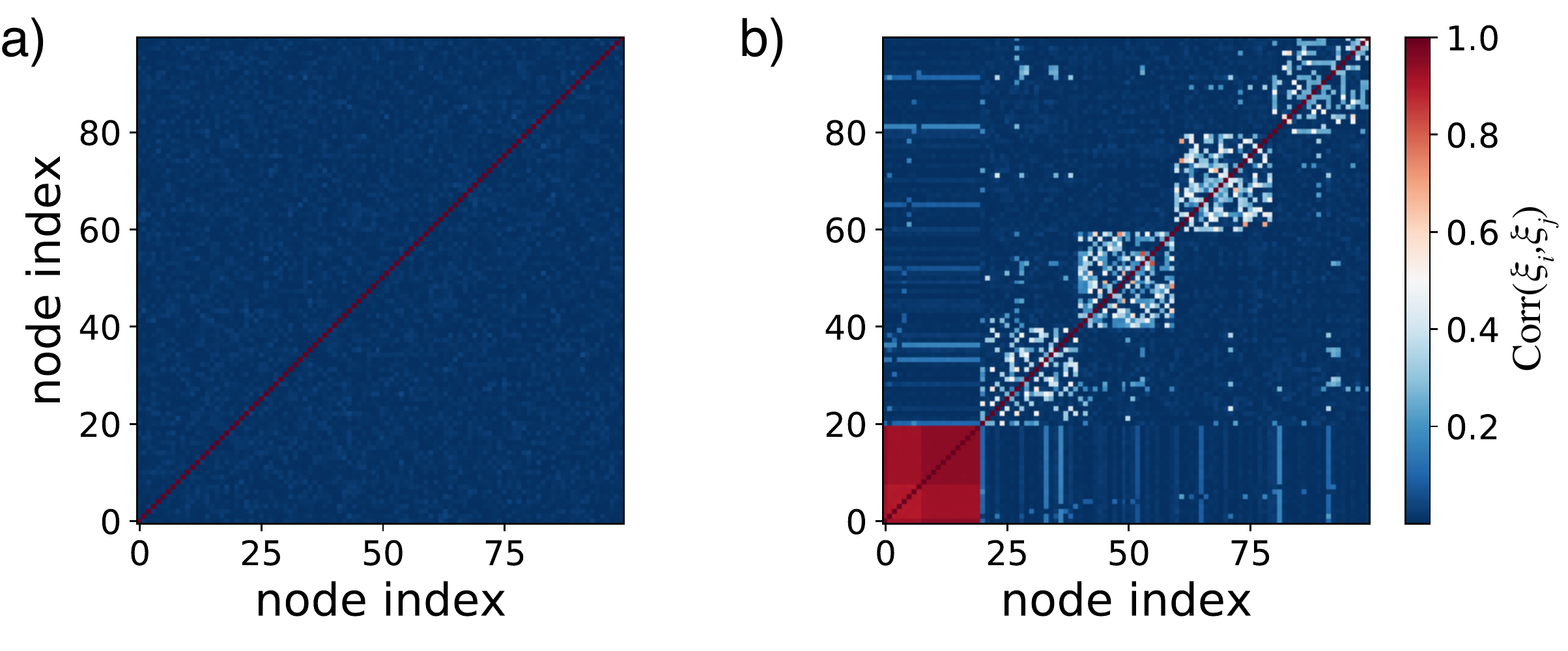}
    \caption{[Color Online] Reconstruction of rich club for spiking neurons with electrical coupling. The left panel shows a color map of the pairwise Pearson-correlation between the time-series of the membrane potential. The right panel, shows the colormap of the noise correlation revealing the community structure and in particular, the integrating cluster. }
\label{spiking}
\end{figure}

Again, we consider the spiking neurons and the electrical coupling described above on the same rich-club motif as before. The reconstruction and analysis of the noise correlation is able to detect the rich-club clusters.  To this end we fix the coupling strength $\alpha=5 \times 10^{-4}$ and simulate the network dynamics. Then we apply our technique to recover the network structure. We show the results in Fig. \ref{spiking}.

\subsection{Bursting neurons coupled with Electric Synapsis on a Rich-Club} 
\begin{figure}[h]
    \centering
        \centering
        \includegraphics[width=0.8\linewidth]{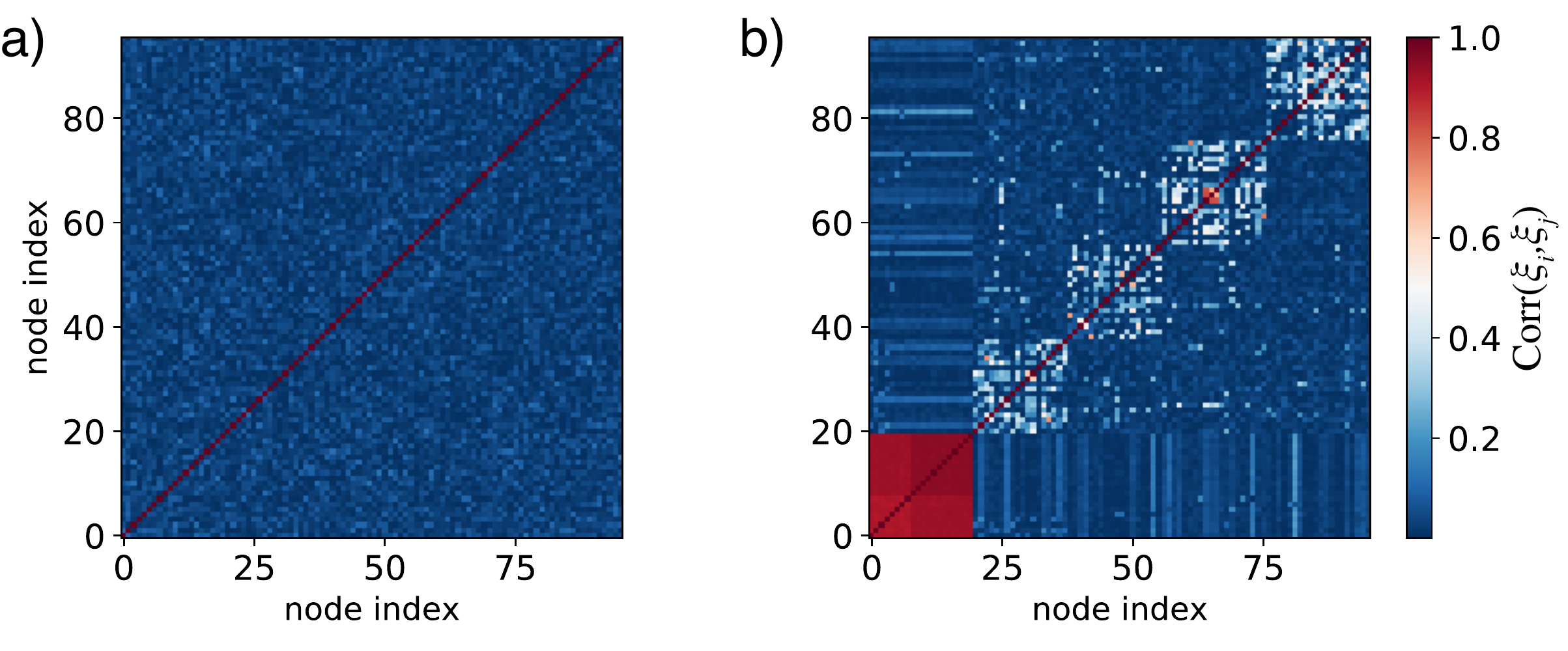}
    \caption{[Color Online] Reconstruction of rich club for bursting neurons with electrical coupling. The left panel shows a color map of the pairwise Pearson-correlation between the time-series of the membrane potential. The right panel, shows the colormap of the noise correlation revealing the community structure and in particular, the integrating cluster.}
\label{bursting}
\end{figure}

Our technique applies equally well then the local dynamics has multiple time-scales such as a bursting neuron. Our extensive numerical investigation reveals that when the resting time is not much larger then the total bursting time the reduced dynamics is capable of extracting the relevant information of the time series. Thus, we fixed the neuron parameter $\beta = 4.4$ to obtain a bursting dynamics. The reconstruction and analysis of the noise correlation is able to detect the rich-club clusters.  To this end we fix the coupling strength $\alpha = 10^{-3}$ and simulate the network dynamics. Then we apply our technique to recover the network structure. We show the results in Fig. \ref{bursting}

\section{Predicting Critical Transitions in Rich-Club Networks}

\begin{figure}[htb]
 \centering
        \begin{subfigure}[t]{\textwidth}
        \centering
        \includegraphics[width=\linewidth]{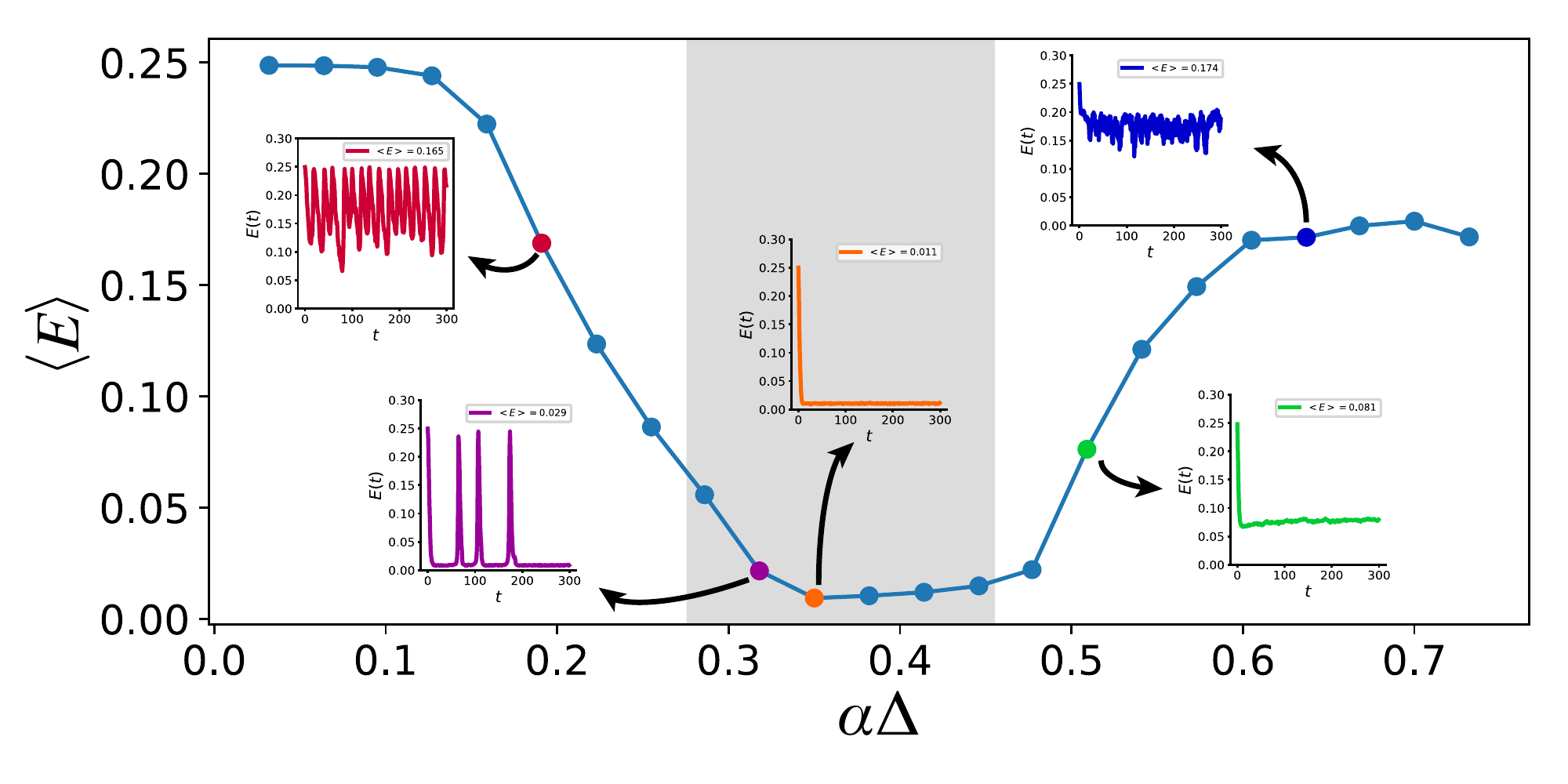}
    \end{subfigure}%
\caption{{\bf Prediction of critical transition in a network with a rich-club motif.} The level of average synchronisation, $\langle E \rangle$ of the integrating cluster is shown for different values of the coupling strength $\alpha$. Insets show $E(t)$ plotted as a function of time for five points indicated by arrows. For $\alpha$ values in the grey shaded region, $\langle E \rangle$ is close to zero and the integrating cluster exhibits collective behaviour. We can predict the extrema of the shaded region by studying the effective network obtained from a time series without any knowledge of parameters, including $\alpha_0 \Delta$. For this prediction, we used the time series when $\alpha_0 \Delta =0.2$ (red point).}
\label{New} 
\vskip -5mm 
\end{figure}

We present another example of how to use the effective network methodology to predict critical transitions. In this case we choose the doubling map for the local dynamics coupled on a rich-club network with clusters sampled as Erd\"os-Renyi graphs. Consider a rich club network of 2200 elements with 5 clusters. Four of the clusters are made of $N_\ell=500$ nodes with small degrees, and one cluster (called integrating cluster or rich club) has $N_{I}=200$ nodes which are connected with most of the network. The edges within a cluster of low degree nodes are  assigned as described above.  
\newline
 As a model for the isolated dynamics, we use the doubling map $f(x)=2x$ mod 1 with the  diffusive coupling $ \bo H(\bo x_i,\bo x_j)=h(x_j,x_i) = \varphi(x_j)-\varphi(x_i)$ where we picked  $\varphi(x)=\sin(2\pi x)$. 
\newline
We use the function
\[
E(t) = \frac{1}{N_I(N_I-1)}\sum_{i, j\in IC} \|x_i(t)-x_j(t)\|
\]
as an empirical measure of the synchronization level at time $t$. From a single multivariate time-series, when the coupling is fixed at $\alpha_0 \Delta = 0.2$ (red marker Figure \ref{New}), the analysis of $ \langle E \rangle$ gives no sign of critical transitions. A  statistical analysis shows that the variance of the averaged synchronization error is not amplified, and the extreme value statistics fails to reveal a transition as it also does an analysis of dynamical correlations. 
\newline
As described above, we can use the effective network methodology to recover the local dynamics $f$ and the effective coupling $v$.  In particular we recover the isolated dynamics $f(x)=2x$ within $4\%$ accuracy for the poorly connected nodes. The unique physical equilibrium measure for this map is the uniform distribution $m$ on $[0,1)$. Furthermore, we recover the effective coupling 
\begin{eqnarray*}
v(x) = \int h(y,x)dm(y)=-\sin 2\pi x.
\end{eqnarray*}
At this point, we know that the approximate evolution of any node $i$ is given by $g_i=f_i+\alpha k_i v_i$, and we can study numerically or analitically this rule to find those values of  $\alpha$ for which the integrating cluster exhibits synchrony.  The analysis shows that excursions towards synchronization will start once the coupling is increased by $15\%$ (see Figure \ref{New}). Below we provide the details on how to recover this critical value of $\alpha$ for which a transition arises.\newline
\newline
{\it Reduction in the Integrating Cluster}. Nodes in the integrating cluster have roughly degree $\Delta$ and make $\kappa \Delta$ connections inside the  integrating cluster and $(1-\kappa)\Delta$ to the rest of the  network.The interactions felt by a node in the rich-club can be split in those coming from nodes in other clusters and those coming from nodes within rich-club itself:
\begin{eqnarray*}
\sum_{j} A_{ij} h(x_i,x_j) =  \sum_{j \in RC} A_{ij} h(x_i,x_j) + \sum_{j \not \in RC} A_{ij} h(x_i,x_j)
\end{eqnarray*}
But,
\begin{eqnarray*}
\sum_{j \not \in RC} A_{ij} h(x_i,x_j) = (1 - \kappa)\Delta \int h(x_i, y) d\mu(y) + \xi^o_i(t)
\end{eqnarray*}
where $\mu$ is the invariant measure for the nodes outside the integrating cluster \footnote{In the example above $\mu$ is the Lebesgue measure $m$.}. Hence, the equation of the integrating cluster can be written as 
\begin{eqnarray*}
x_i(t+1) = q_i(x_i(t)) +  \sum_{j \in RC} A_{ij} h(x_i,x_j) + \xi^o_i(t),
\end{eqnarray*}
where $q_i(x_i(t)) = f_i(x_i(t)) + (1 - \kappa)\Delta\alpha \int h(x_i, y) d\mu(y)$.
\newline
We can estimate $\mu$ empirically analyzing each cluster. Assuming that $h(x,y)=\varphi(y) -\varphi(x)$,  we can recover $\varphi$ from the analysis of $q_i$ using the reconstruction techniques. In fact, for this particular choice of $h$, $\int h(x,y) d\mu(y)$ is equal to $-\varphi(x)$ plus a constant. Now if $\nu$ is the measure that describes the behaviour of a node in the integrating cluster, then the interaction within the rich-club can be written as
\begin{equation}
\sum_{j \in RC} A_{ij} h(x_i,x_j)=\kappa\Delta\int  h(x_i, y) d\nu(y)+\xi^c(t)=\kappa \Delta\left(-\varphi(x)+\int \varphi(s)d\nu(s)\right)+\xi^c(t).
\end{equation}
Putting the two equations together one has that 
\begin{equation}\label{Eq:nuself}
x_i(t+1) =  g_i(x_i(t)) + \alpha \Delta c(\mu,\nu) + \zeta_i(t)
\end{equation}
where $g_i = 
f_i-\alpha \Delta \varphi $ models the reduced dynamics and the $c(\mu,\nu) = (1 - \kappa)\Delta\alpha\int\varphi (s)d\mu(s)+\kappa\Delta\alpha\int \varphi(s)d\nu(s)$ is the mean contribution from all interactions (inside and outside the integrating cluster). Finally $\zeta_i =  \xi^o_i(t)+\xi^c(t)$ combines the effect of the fluctuations.  
\newline
{\it Model from Data.} From a single multivariate time series at a given coupling parameter, shown in Figure \ref{New} as a red dot,  we reconstruct the model. First, we obtain the  the rule $g_i$ which we uncover to be $g_i(x) = 2 x- \beta_i \sin 2\pi x $ mod $1$, hence $\beta_i = 0.169$ and this number is nearly independent of the node in the integrating cluster. Hence, we obtain an estimate $(\alpha \Delta)_{\rm est} = 0.169$. We also obtain that the dynamics of nodes in the communities is well approximated by $f(x) = 2x $ mod $1$. 
\newline
Next,  we need to estimate $\kappa$ to construct a model for the connectivity of the integrating cluster.  From the recovered local dynamics $f$, one knows that  $\mu$ is the Lebesgue measure and, since $\int \varphi d\mu = 0$, we obtain $ c(\mu,\nu) = \kappa \int \varphi d\nu$. 
In the regime of parameters where the measurements have been made, $\nu$ can be obtained empirically and this allows to recover $\kappa$ since
\[
\frac{1}{\int \varphi d\nu} \frac{1}{\alpha \Delta} \langle x_i(t+1) - g_i(x_i(t))  \rangle \rightarrow  \kappa,
\]
where we evaluate $\int\phi d\nu$ with respect to the empirically retrieved $\nu$, we substitute $\alpha\Delta$ with the estimated value above, and  $\langle \cdot \rangle$ denotes the time average. We obtain that $\kappa = 0.47$. The data analysis reveals that such $\kappa$ value is nearly independent of the node in the integrating cluster. 
Moreover, the analysis of the covariance of the noise in the integrating clusters shows a lack of communities and the functional analysis network indicates that the integrating cluster of  is a random network 200 nodes with $p=0.83$. From this analysis, we obtain an estimate for $\Delta \approx p \times 200 / \kappa =  353=\Delta_{est}$. We are now able to estimate $\alpha$ by $\alpha_{est} = (\alpha\Delta)_{est}/\Delta_{est}= 5\times 10^{-4}$.
\newline
{\it Synchronization Prediction.} With the reconstructed data $(f,v,A)$, we obtain that  for a node in the integrating cluster, \eqref{Eq:nuself} reads as
\[
x_i(t+1) =  2x_i(t)-\alpha \Delta\sin(2\pi x_i(t)) +\kappa\Delta\alpha \int \sin(2\pi s)d\nu(s)+ \xi(t).
\]
When $1/2\pi<\alpha\Delta<3/2\pi$, the map $  2x_i(t)-\alpha \Delta \sin(2\pi x_i(t))$ mod $1$ has an attracting fixed point at 0. In this new regime, $\nu=\delta_0$ is a self-consistent measure for the  integrating cluster, meaning that the measure $\nu$ gives rise to an approximated dynamical rule $g_i$ that has $\nu$ as equilibrium measure.  This can be easily verified since $\int \sin(2\pi s)d\delta_0(s)=0$. We can then conclude that by selecting a value of the coupling strength such that $\alpha \Delta$ is in the range above, one expects all the states at the nodes in the integrating cluster to evolve towards the point $0$ and fluctuate around this point by $\xi(t)$. 
\newline
To obtain the range of $\alpha \Delta$ such that the fluctuations of $E$ are around twice $\max\| \xi  \|$, we notice that since $\nu=\delta_0$ the stability properties are given by the linear stability around $x=0$. Let $u$ be a small displacement around $0$ and let us denote at  $J(\alpha) = Dg_i(0)$ the Jacobian of the map $g_i$ at $0$. Then we obtain that $u(t+1) = \sum_{i=0}^t J(\alpha)^{i} \xi_i$ and so $\| u \| \le \max \| \xi \| / (1 - J(\alpha))$, which yields that $1/(1-J(\alpha)) < 2$. Hence, we obtain $3/4\pi<\alpha\Delta<5/4\pi$. Because we measured  $\alpha \Delta$ as $0.169$ in the data given we predict that a high quality coherent state in the rich-clubwill appear then  $\alpha \Delta$ is increased by  $40\%$. This is a agreement with the experiments. 

\section{Cat cerebral cortex}

\subsection{Bursting neurons with Chemical Synapesis}

\begin{figure}[h]
    \centering
        \centering
        \includegraphics[width=1.0\linewidth]{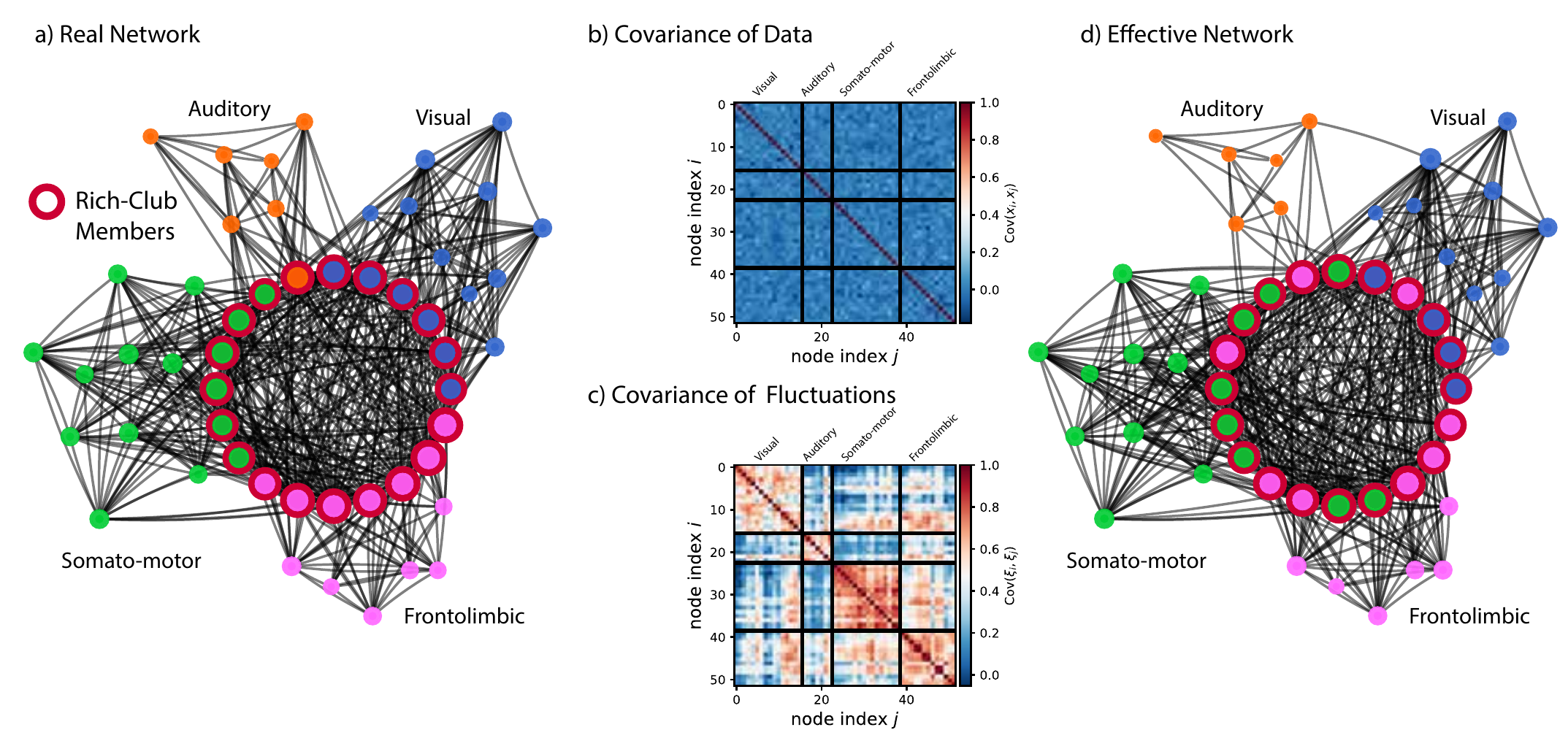}
    \caption{Effective network of the cat cerebral cortex. We use the local dynamics as a spiking neuron coupled via electric synapses with the parameters $\Delta \alpha=0.05$ and $\beta=4.4$. (a) The cat cerebral cortex network with nodes colour coded according to the four functional modules. Rich-club members are indicated by red encircled nodes. (b) The covariance matrix of the data cannot detect communities. (c) The covariance matrix of the fluctuations can distinguish clusters of interconnected nodes. (d) A model in the cat cortex constructed via the effective network approach. From the matrix in (c) we can recover a representative effective network. The reconstructed network represents the real network in (a) with good accuracy. See Methods for the details of the detection of communities and rich-club members.}
\label{catmap_bursting}
\end{figure}

We simulated each mesoregion of the cat cerebral cortex network with bursting Rulkov oscillators coupled through chemical synapsis. 
For such dynamics the parameters are given as $\beta=4.4$ and $\Delta \alpha=0.05$ where there is no synchrony between oscillators Fig.~\ref{catmap_bursting}. We use the simulated data to reconstruct the network structure as shown in Figure \ref{catmap_bursting}.

\begin{thebibliography}{99}

\bibitem{kandel2000} Kandel, E.R., Schwartz, J.H. \& Jessell, T.M. eds.,  {\it Principles of neural science}, Vol. 4. (New York: McGraw-hill, 2000). 

\bibitem{bohland2009} Bohland, J. W. {\it et al.} A proposal for a coordinated effort for the determination of brainwide neuroanatomical connectivity in model organisms at a mesoscopic scale. {\it PLoS~Computational~Biology} {\bf 5}, e1000334 (2009).

\bibitem{de2004discovery} De La Fuente, A., Bing, N., Hoeschele, I., \& Mendes, P.  Discovery of meaningful associations in genomic data using partial correlation coefficients. {\it Bioinformatics}, {\bf 20(18)}, 3565-3574 (2004).

\bibitem{reverter2008combining} Reverter, A., \& Chan, E. K.  Combining partial correlation and an information theory approach to the reversed engineering of gene co-expression networks. {\it Bioinformatics}, {\bf 24(21)}, 2491-2497 (2008).

\bibitem{butte1999mutual} Butte, A. J., \& Kohane, I. S.  Mutual information relevance networks: functional genomic clustering using pairwise entropy measurements. {\it In Biocomputing 2000} (pp. 418-429), (1999).

\bibitem{braunstein2008inference} Braunstein, A., Pagnani, A., Weigt, M., \& Zecchina, R. Inference algorithms for gene networks: a statistical mechanics analysis. {\it Journal of Statistical Mechanics: Theory and Experiment}, {\bf 2008(12)}, P12001 (2008).

\bibitem{cocco2009neuronal} Cocco, S., Leibler, S., \& Monasson, R.  Neuronal couplings between retinal ganglion cells inferred by efficient inverse statistical physics methods. {\it Proc.~Natl.~Acad.~Sci.~USA}, {\bf 106}(33), 14058-14062 (2009).

\bibitem{bressler2011wiener} Bressler, S. L., \& Seth, A. K.  Wiener–Granger causality: a well established methodology. {\it Neuroimage}, {\bf 58}(2), 323-329 (2011).

\bibitem{ladroue2009beyond} Ladroue, C., Guo, S., Kendrick, K., \& Feng, J. Beyond element-wise interactions: identifying complex interactions in biological processes. {\it PloS one}, {\bf 4}(9), e6899 (2009)

\bibitem{wang2016} Wang, W., Lai, Y., \& Grebogi, C. Data based identification and prediction of nonlinear and complex dynamical systems. {\it Physics Reports}, {\bf 644}, 1--76 (2016).

\bibitem{casadiego2017} Casadiego, J., Nitzan, M., Hallerberg, S., \& Timme, M. Model-free inference of direct network interactions from nonlinear collective dynamics. {\it Nat.~Commun.} {\bf 8}, 2192 (2017).

\bibitem{han2015} Han, X., Shen, Z., Wang, W. X., \& Di, Z..  Robust reconstruction of complex networks from sparse data. {\it Phys.~Rev.~Lett.}, {\bf114}, 028701 (2015).

\bibitem{nitzan2017revealing} Nitzan, M., Casadiego, J., \& Timme, M.  Revealing physical interaction networks from statistics of collective dynamics. {\it Science~advances}, {\bf 3(2)}, e1600396 (2017).

\bibitem{stankovski2017} Stankovski, T., Pereira, T., McClintock, P. V., \& Stefanovska, A.  Coupling functions: universal insights into dynamical interaction mechanisms. {\it Reviews of Modern Physics}, {\bf 89}(4), 045001 (2017).

\bibitem{schneidman2006} Schneidman, E., Berry, M. J., Segev, R., \& Bialek, W.  Weak pairwise correlations imply strongly correlated network states in a neural population. {\it Nature} {\bf 440}, 1007--1012 (2006).

\bibitem{haas2015} Haas, J. S.  A new measure for the strength of electrical synapses. {\it Front.~Cell.~Neurosci.}, {\bf 9}, 378 (2015).

\bibitem{heuvel2011} Van Den Heuvel, M. P., \& Sporns, O.  Rich-club organization of the human connectome. {\it J.~Neurosci.}, {\bf 31}, 15775--15786. (2011).

\bibitem{park2013} Park, H. J., \& Friston, K. Structural and functional brain networks: From connections to cognition. {\it Science} {\bf 342}, 6158 (2013).

\bibitem{pereira2017} Pereira, T., van Strien, S., \& Tanzi, M. Heterogeneously coupled maps: hub dynamics and emergence across connectivity layers. {\it To appear in J.
Eur. Math. Soc., preprint: arXiv} {\bf 1704.06163}, 1--63 (2017)

\bibitem{izhikevich2007} Izhikevich, E. M. {\it Dynamical systems in neuroscience}, (MIT press, 2007).

\bibitem{yadav2017} Yadav, P., McCann, J. A., \& Pereira, T. Self-synchronization in duty-cycled internet of things (IoT) applications. {\it IEEE~Internet~of~Things~Journal} {\bf 4}, 2058--2069 (2017).

\bibitem{dorfler2013} D\"orfler, F., Chertkov, M., \& Bullo, F.  Synchronization in complex oscillator networks and smart grids. {\it Proc.~Natl.~Acad.~Sci.~USA} {\bf 110}, 2005--2010 (2013).

\bibitem{watanabe1994} Watanabe, S., \& Strogatz, S. H. Constants of motion for superconducting Josephson arrays. {\it Physica~D:~Nonlinear~Phenomena} {\bf 74}, 197--253 (1994).


\bibitem{winfree2001} Winfree, A.T.  {\it The Geometry of Biological Time}, Interdisciplinary Applied Mathematics: Vol 12 (Springer-Verlag~New~York, 2001).

\bibitem{pinto2000} Pinto, R. D., Varona, P., Volkovskii, A. R., Sz\"ucs, A., Abarbanel, H. D. I., \& Rabinovich, M. I.  Synchronous behavior of two coupled electronic neurons. {\it Phys.~Rev.~E} {\bf 62}, 2644--2656 (2000).

\bibitem{eroglu2017} Eroglu, D., Lamb, J. S. W., \& Pereira, T. Synchronisation of chaos and its applications. {\it Contemporary Physics} {\bf 58}, 207--243 (2017). 

\bibitem{scannell1993}Scannell, J. W. \& Young, M. P.  The connectional organization of neural
systems in the cat cerebral cortex. {\it Curr. Biol.} {\bf 3}, 191--200 (1993).

\bibitem{scannell1995} Scannell, J. W., Blakemore, C. \& Young, M. P.  Analysis of connectivity
in the cat cerebral cortex. {\it J. Neurosci.} {\bf 15}, 1463--1483 (1995).

\bibitem{zamora-lopez2010} Zamora-L\'opez, G., Zhou, C., \& Kurths, J.  Cortical hubs form a module for multisensory integration on top of the hierarchy of cortical networks. {\it Frontiers in Neuroinformatics} {\bf 4}, 1--13 (2010).

\bibitem{takemura2013} Takemura, S., {\it et al.} A visual motion detection circuit suggested by Drosophila connectomics. {\it Nature} {\bf 500}, 175--181 (2013).

\bibitem{garcia-perez2018} Garc\'ia-P\'erez, G., Bogu\~n\'a, M., \& Serrano, M. \'A. Multiscale unfolding of real networks by geometric renormalization. {\it Nat.~Phys.} (2018).

\bibitem{shandilya2011} Shandilya, S. G., \& Timme, M. Inferring network topology from complex dynamics. {\it New Journal of Physics}, {\bf 13}, 013004 (2011). 


\bibitem{james2013} James, G., Witten, D., Hastie, T., \& Tibshirani, R. {\it An introduction to statistical learning}, Vol. 112. (New York: Springer, 2013.)

\bibitem{brunton2015} Brunton, S. L., Proctor, J. L., \& Kutz, J. N. Discovering governing equations from data: Sparse identification of nonlinear dynamical systems. {\it Proc.~Natl.~Acad.~Sci.~USA} {\bf 113}, 3932--3937 (2015).

\bibitem{mangan2016} Mangan, N. M., Brunton, S. L., Proctor, J. L., \& Kutz, J. N.  Inferring Biological Networks by Sparse Identification of Nonlinear Dynamics. {\it IEEE Trans on Molecular, Biological, and Multi-Scale Communications}, {\bf 2}, 52--63 (2016).

\bibitem{wang2011} Wang, W. X., Yang, R., Lai, Y. C., Kovanis, V., \& Grebogi, C. Predicting catastrophes in nonlinear dynamical systems by compressive sensing.  {\it Phys.~Rev.~Lett.} {\bf 106}, 154101 (2011). 



\bibitem{eguiluz2005} Eguiluz, V. M., Chialvo, D. R., Cecchi, G. A., Baliki, M., \& Apkarian, A. V. Scale-free brain functional networks. {\it Phys.~Rev.~Lett.} {\bf 94}, 018102 (2005)

\bibitem{bullmore2009} Bullmore, E., \& Sporns, O. Complex brain networks: Graph theoretical analysis of structural and functional systems. {\it Nat.~Rev.~Neuroscience} {\bf 10}, 186--198 (2009).



\bibitem{bettinardi2017} Bettinardi, R. G., {\it et al.} How structure sculpts function: Unveiling the contribution of anatomical connectivity to the brain's spontaneous correlation structure. {\it Chaos} {\bf 27}, 047409  (2017).

\bibitem{zhang2006} Zhang, J., \& Small, M.,  Complex network from pseudoperiodic time series: Topology versus dynamics.   {\it Phys.~Rev.~Lett.} {\bf 96}, 238701 (2006).

\bibitem{greicius2003} Greicius, M. D., Krasnow, B., Reiss, A. L., \& Menon, V.  Functional connectivity in the resting brain: a network analysis of the default mode hypothesis. {\it Proc.~Natl.~Acad.~Sci.~USA} {\bf 100}, 253--258 (2003).


\bibitem{colizza2006} Colizza, V., Flammini, A., Serrano, M. A., \& Vespignani, A.  Detecting rich-club ordering in complex networks. {\it Nat. Phys.} {\bf 2}, 110 (2006).

\bibitem{blondel2008} Blondel, V. D., Guillaume, J.-L., Lambiotte, R. \& \'Etienne, L. Fast unfolding
of communities in large networks. {\it J. Stat. Mech.} {\bf 2008}, P10008 (2008).

\bibitem{lopes2017} Lopes, M. A., Richardson, M. P., Abela, E., Rummel, C., Schindler, K., Goodfellow, M., \& Terry, J. R. An optimal strategy for epilepsy surgery: Disruption of the rich-club? {\it PLoS computational biology} {\bf 13}, e1005637 (2017).


\bibitem{rulkov2001} Rulkov, N. F.  Regularization of synchronized chaotic bursts. {\it Phys.~Rev.~Lett.} {\bf 86}, 183--186 (2001).



\bibitem{pereira2010}  Pereira, T. Hub synchronization in scale-free networks. {\it Phys.~Rev.~E}, {\bf 82}(3), 036201(2010).

\bibitem{li2017universal} Li, Y., et al. Universal style transfer via feature transforms. {\it Advances in neural information processing systems}, pp. 386-396, (2017).


%
%
%
%
%

\bibitem{chung2002} Chung, F. \& Lu, L. Connected components in random graphs with given expected degree sequences. {\it Annals of Combinatorics} {\bf 6}, 125 (2002).

\bibitem{chung2003} Chung, F., Lu, L. \& Vu, V. Eigenvalues of Random Power law Graphs. {\it Annals of Combinatorics} {\bf 7}, 21 (2003).

\bibitem{muniruzzaman1957} Muniruzzaman, A. N. M.  On Measures of location and dispersion and tests of hypotheses in a pare to population. {\it Calcutta Statistical Association Bulletin} {\bf 7}, 115--123 (1957).

\bibitem{hill1975} Hill, B. M.  A Simple general approach to inference about the tail of a distribution. {\it Ann. Statist.} {\bf 3}, 1163--1174 (1975).


\bibitem{clauset2009} Clauset, A., Shalizi, C. R. \& Newman M. E. J.  Power-Law distributions in empirical data. {\it SIAM Review} {\bf 51}, 661--703 (2009).

\bibitem{newman2003} Newman, M.E.J. The structure and function of complex networks. {\it SIAM Review} {\bf45} 167--256, (2003)

\bibitem{footnotesmoothing} Cleveland, W.S. Robust Locally Weighted Regression and Smoothing Scatterplots. {\it Journal of the American Statistical Association} {\bf 74} 368:829-836 (1979).

\bibitem{pereira2007} Pereira, T., M. S. Baptista, \& J. Kurths. Phase and average period of chaotic oscillators. {\it Physics Letters A } {\bf 362}, 159-165 (2007).




%
%


%
%
%
































%













 














\end{thebibliography}

\begin{thebibliography}{99}

\bibitem{Tanzi} Pereira, T., van Strien, S., \& Tanzi, M. {\it Heterogeneously coupled maps: hub dynamics and emergence across connectivity layers}. J.
Eur. Math. Soc., preprint: arXiv  {\bf 1704.06163}, 1--63 (2017)

\bibitem{Strien} de Melo, W., and Van Strien, S., {\it One-dimensional dynamics}. Springer (2012).

\bibitem{Roessler} R\"ossler, O. E. {\it An equation for continuous chaos}. Physics Letters A 57, no. 5 (1976): 397-398.


\end{thebibliography}
\end{document}